\documentclass[11pt,a4paper]{article}
\usepackage{jheppub}

\usepackage{color}
\usepackage[dvipsnames, svgnames, x11names]{xcolor}
\usepackage{lmodern}

\usepackage[utf8]{inputenc}
\usepackage{float}
\usepackage{graphicx}
\usepackage{setspace}
\usepackage{subfigure}
\usepackage{bm}
\usepackage{bbm}
\usepackage{amsmath}
\usepackage{amssymb}
\usepackage{amsbsy}
\usepackage{amsthm}
\usepackage{amsfonts}
\usepackage{braket}
\usepackage{multirow}
\usepackage{slashed}
\usepackage{babel}
 \hypersetup{
	colorlinks=true,
	linkcolor=cyan,
	urlcolor=blue,
	citecolor=red,
}

\def\nn{\nonumber}
\def\Tr{\text{Tr}}
\def\sgn{\text{sgn}}

\def\L{\mathcal{L}}

\def\H{\mathcal{H}}
\def\S{\mathcal{S}}

\def\avg#1{\left\langle#1\right\rangle}
\def\bra#1{\left\langle#1\right|}
\def\ket#1{\left|#1\right\rangle}

\def\kc#1{\left(#1\right)}
\def\kd#1{\left[#1\right]}
\def\ke#1{\left\{#1\right\}}

\def\sgn{{\rm sgn}}

\def\be{\begin{equation}}       \def\ee{\end{equation}}
\def\bea{\begin{eqnarray}}      \def\eea{\end{eqnarray}}
\def\ba{\begin{array}}
	\def\ea{\end{array}}
\def\bnum{\begin{enumerate} }
	\def\enum{\end{enumerate}}

\def\nn{\nonumber}

\def\=>{\Rightarrow}
\def\>{\rightarrow}

\def\eye2{Fathbb{I}}

\def\Tr{\mathrm{Tr}}

\title{Entanglement inside a black hole before the Page time}

\author[a]{Yuxuan Liu,}
\emailAdd{223093@csu.edu.cn}
\affiliation[a]{Institute of Quantum Physics, School of Physics, Central South University, Changsha 418003, China}

\author[b]{Shao-Kai Jian,}
\emailAdd{sjian@tulane.edu}
\affiliation[b]{Department of Physics and Engineering Physics, Tulane University, \\New Orleans, Louisiana, 70118, USA}

\author[c]{Yi Ling,}
\emailAdd{lingy@ihep.ac.cn}
\affiliation[c]{Institute of High Energy Physics, Chinese Academy of Sciences, Beijing 100049, China\\ School of Physics, University of Chinese Academy of Sciences,
Beijing 100049, China}

\author[d,*]{and Zhuo-Yu Xian\note[*]{Corresponding author.}}
\emailAdd{zhuo-yu.xian@physik.uni-wuerzburg.de}
\affiliation[d]{Institute for Theoretical Physics and Astrophysics and W\"urzburg-Dresden Cluster of Excellence ct.qmat, Julius-Maximilians-Universit\"at W\"urzburg, 97074 W\"urzburg, Germany}

\abstract{
We investigate the evolution of entanglement within an open, strongly coupled system interacting with a heat bath as its environment, in the frameworks of both the doubly holographic model and the Sachdev-Ye-Kitaev (SYK) model. 
Generally, the entanglement within the system initially increases due to internal interactions; however, it eventually dissipates into the environment.
In the doubly holographic setup, we consider an end-of-the-world brane in the bulk to represent an eternal black hole coupled with its radiation and the evolution of the global thermofield double (TFD) state.
For small black holes, the reflected entropy between the bipartition exhibits a ramp-plateau-slump behavior, where the plateau arises due to the phase transition of the entanglement wedge cross-section before the Page time. Similarly, the mutual information between the bipartition displays a ramp-slop-stabilizing behavior. In quantum mechanics, we consider a double copy of the SYK-plus-bath system in a global TFD state, resembling an eternal black hole interacting with an environment. 
The R\'enyi mutual information within the double-copied SYK clusters exhibits a ramp-plateau-slope-stabilizing behavior. 
The dynamic behaviors of the entanglement quantities observed in these two models are attributable to the competition between the internal interaction of the system and the external interaction with the baths. 
Our study provides a fine-grained picture of the entanglement dynamics inside black holes before their Page time.
}

\arxivnumber{}

\begin{document}
	
\maketitle

\section{Introduction}
Understanding the nature of Hawking radiation in black hole physics requires us to consider the evaporation procedure from the quantum mechanical point of view, which is a formidable challenge due to the intricacy of the interaction between quantum fields and spacetime itself. This challenge leads to the renowned black hole information paradox, which asks whether black hole evaporation is a unitary process \cite{hawking1974black,hawking1975particle,hawking1976breakdown,susskind1993stretched,harlow2013quantum,Almheiri:2012rt,Maldacena:2013xja}. 
The early work by Hawking suggests that the nearly thermal spectrum of Hawking radiation implies a continuous increase in the entanglement between the black hole and the radiation, such that the information would be unavoidably lost after the black hole has evaporated completely. 
However, from the perspective of unitary quantum physics, in particular, viewing the black hole and the radiation as a joint system described by quantum states which are governed by random evolution, the entanglement entropy of the radiation is expected to decrease at late stages, aligning with the dynamics depicted by the Page curve \cite{Page:1993wv,Page:2004xp,Page:2013dx}.

Recently, the Gauge/Gravity duality offers a clear correspondence between black holes in AdS space and strongly coupled quantum systems, where unitarity could be implemented during evolution. 
Within this framework, significant progress has been made in revealing the entropy dynamics during black hole evaporation \cite{Penington:2019npb,Almheiri:2019psf}.
This advancement can be ascribed to the fact that the generalized gravitational entropy can be geometrically described by extremizing the area of the minimal surface \cite{Lewkowycz:2013nqa,Engelhardt:2014gca}, allowing the entanglement attributed to Hawking radiation to be encapsulated by the quantum extremal surface (QES), as delineated by the island formula \cite{Almheiri:2019hni} 
\begin{equation}\label{eq:QESinRad}
S[\mathcal{R}]=\min_{\mathcal I} \left\{ \mathop{\text{ext}}\limits_{\mathcal I} \left[\frac{\textbf{A}[\partial \mathcal{I}]}{4 G_{N}} + S[\mathcal{R} \cup \mathcal{I}]\right]\right\},
\end{equation}
with $\textbf A$ the area, $\mathcal{R}$ and $\mathcal{I}$ the radiation and the islands, respectively.
According to the island formula (\ref{eq:QESinRad}), the entanglement entropy of radiation can be computed, and it turns out that throughout the evolution it aligns with the Page curve \cite{Almheiri:2020cfm}.
See \cite{Alishahiha:2020qza,Hashimoto:2020cas,Anegawa:2020ezn,Hartman:2020swn,Chen:2020jvn,Bhattacharya:2020uun,Deng:2020ent,Wang:2021woy,He:2021mst,Gautason:2020tmk,Krishnan:2020oun,Sybesma:2020fxg,Chou:2021boq,Hollowood:2021lsw,Suzuki:2022xwv,Suzuki:2022yru,Bhattacharya:2021nqj,Bhattacharya:2021dnd,Caceres:2021fuw,Bhattacharya:2021jrn,Caceres:2020jcn,Chen:2019iro,Balasubramanian:2020hfs,Anderson:2020vwi,Balasubramanian:2021xcm,Engelhardt:2022qts,Chou:2023adi,Jeong:2023lkc,Ahn:2021chg,Karch:2022rvr,Uhlemann:2021nhu} for further investigations and discussions on the island paradigm. 
This formula was initially introduced in a model featuring two-dimensional gravity coupled with two-dimensional CFT matter sectors \cite{Almheiri:2019hni}. 

In addition to black hole evaporation, the information paradox also manifests in the equilibrating process between an eternal black hole and its thermal radiation, maintained at the same temperature \cite{Almheiri:2019yqk,Penington:2019kki}.
Despite the absence of energy exchange in the radiation exchange process, the entanglement between the black hole and the radiation does not grow indefinitely. 
Consequently, a similar Page curve is expected to emerge, indicating a turnover in entanglement entropy over time. 
Notably, the concept of the island remains relevant and applicable in these static geometries.

To compute the second term in the island formula (\ref{eq:QESinRad}) in a non-perturbative way, one can consider the radiation as a $2$-dimensional CFT \cite{Almheiri:2019yqk} or a holographic CFT, which allows for a higher dimensional gravity dual, namely doubly holographic setup \cite{Almheiri:2019hni,Chen:2019uhq,Almheiri:2019psy}, generalized from the holographic dual of boundary conformal field theory (BCFT) \cite{Takayanagi:2011zk,Nozaki:2012qd,Izumi:2022opi,Erdmenger:2014xya,Erdmenger:2015spo}. 
In the doubly holographic setup, the black hole is modeled by a tensioned brane under Neumann boundary condition \cite{Chu:2018ntx,Miao:2018qkc} and the QES is translated into a standard Hubeny-Rangamani-Takayanagi (HRT) surface within the higher dimensional geometry.
In this framework, realizing a dynamical Page curve requires a sufficient number of degrees of freedom (DOF) on the brane \cite{Ling:2020laa,Chen:2020uac,Geng:2020fxl,Geng:2020qvw}, and various methods for incorporating adequate DOF have been suggested \cite{Ling:2020laa}. 
For further explorations of brane-world models in higher dimensions, refer to \cite{Krishnan:2020fer,Geng:2020fxl,Geng:2020qvw,Chen:2020uac,Chen:2020hmv,Hernandez:2020nem,Grimaldi:2022suv,Miao:2020oey,Akal:2020wfl,Akal:2020twv,Omidi:2021opl}.
In path integral formalism, islands arise due to the predominance of a specific saddle point, which serves to link different replicas and is called replica wormholes \cite{Penington:2019kki,Almheiri:2019qdq,Rozali:2019day,Karlsson:2020uga}. 
It becomes dominant when the black hole has been coupled to the radiation for a long time, which leads to the Page time.

From the boundary point of view, the above equilibrating process corresponds to the evolution of the system-plus-bath in a global thermofield-double (TFD) state. 
The interaction between the system and the bath leads to the production of entanglement entropy without energy exchange. 
However, due to the finite size effect of the system, the entanglement entropy ceases to grow and saturates to the thermal entropy after the final thermalization of the system in its environment.
Microscopic models could be realized in the SYK clusters \cite{Kitaev:2015a,Maldacena:2016remarks} coupled to heat baths, where the baths could be taken as other SYK clusters \cite{Zhang:2019evaporation,Penington:2019kki,Almheiri:2019jqq,Gu:2016oyy,Gu:2017njx,Jian:2020krd} or Majorana chains \cite{Chen:2020wiq,Liu:2022pan}. 
In addition to the replica diagonal solution \cite{Gu:2017njx}, a new replica wormhole solution emerges. 
This configuration demonstrates a robust correlation among distinct replicas, mediated by the intersecting interaction between different replicas of the systems and the baths. 
It becomes dominant after the Page time and leads to the saturation of entropy. 

However, to completely understand the dynamic behavior of the entanglement in the whole system during evolution, the mere study of the entanglement between radiation and black holes is not enough. 
Instead, one needs to further consider the interior entanglement inside the black hole. 
It is well known that the environment renders the decoherence and disentanglement within a system \cite{PhysRevA.69.052105,PhysRevA.75.062312,kim2012protecting,DelCampo:2019afl,Yang:2024boo}, where interactions between the system and environment can lead to a reduction of system purity and internal entanglement. 
Such effects are anticipated during the evolution of radiating black holes and SYK clusters coupled to baths. Nevertheless, the chaotic nature of black holes and the SYK model, which promotes rapid entanglement growth within the system, must also be taken into account. 
Consequently, the evolution of interior entanglement is likely a result of competing dynamics: the thermalization driven by interactions with the bath and the local thermalization stemmed from the internal interactions within the system.

We anticipate a common occurrence of competitive dynamics during the evolution of radiating black holes and strongly coupled open systems. 
The evolution of entanglement between separate black holes has been explored in previous studies \cite{Liu:2022pan,Afrasiar:2022ebi,Geng:2021iyq,Geng:2021mic}, as well as within the SYK model \cite{Wang:2023vkq}. 
However, these scenarios lack direct strong interactions between distinct black holes or between two SYK clusters. 
Therefore, the type of competition in the literature is almost one-sided, primarily driven by the interaction of the system with its environment rather than interactions within the individual components of the system. 
Consequently, our study aims to further explore the entanglement properties within strongly coupled systems, by focusing on the interior of gravity or SYK systems.
We will particularly focus on how internal interactions generate entanglement within the system, and how interactions with the environment (typically a heat bath) can disrupt this internal entanglement. 
This paper will analyze such competing dynamics in the following two strongly coupled systems:
Firstly, we set up a finite-sized gravity system coupled to a heat bath, and calculate the interior entanglement within this gravity system. 
Secondly, we construct a combined SYK system by coupling a SYK cluster to a huge SYK bath, and in a similar vein, we focus on the interior entanglement within the SYK cluster. 
For simplicity, in the gravitational system, the interior entanglement will be measured by the reflected entropy, which was applied to the black hole information problem \cite{Chandrasekaran:2020qtn,Li:2021dmf,Li:2020ceg,Czech:2023rbh}.
While in the SYK system, it will be measured by the R\'enyi mutual information.

The structure of our paper is organized as follows: in Section \ref{sec:gravity}, we explore a doubly holographic setup in $(3+1)$ dimensions, which corresponds to a black hole on the brane coupled to a heat bath at its boundary. 
We then compute the holographic reflected entropy and mutual information respectively, to describe the mixed-state entanglement of the black hole bipartition.
In Section \ref{sec:SYK}, we couple the SYK clusters to baths and calculate the second R\'enyi mutual information between the bi-partition of the SYK clusters. 
By numerically solving the Schwinger-Dyson equation, we find replica wormhole solutions and observe a growth-decay behavior for the second R\'enyi mutual information before the Page time. 
Conclusions and discussions are presented in Section \ref{sec:conc}.


\section{The doubly holographic setup}\label{sec:gravity}
Within the framework of double holography \cite{Almheiri:2019hni,Chen:2020uac}, we consider a $(d+1)$-dimensional AAdS spacetime, which is intersected by a $d$-dimensional end-of-the-world (EOW) brane, denoted as $\mathcal{B}$. This brane meets the conformal boundary $\bm{\partial}$ at an intersection of dimension $(d-1)$, which we denote by $\partial\mathcal{B}$ \cite{Almheiri:2019hni,Chen:2020uac,Almheiri:2019yqk,Takayanagi:2011zk}. This system can be equivalently described from three different points of view as follows.
Firstly, we articulate the action of the gravity theory in $(d+1)$ dimensions as:
\begin{align}\label{eq:Action}
I=
&\frac{1}{16\pi G_N^{(d+1)}} \Bigg[ \int d^{d+1}x
\sqrt{-g}\left(R+\frac{d(d-1)}{L^2}\right)+2\int_{\bm{\partial}}d^{d}x\sqrt{-h_{\bm{\partial}}}K_{\bm{\partial}}\nonumber\\
& + 2 \left(\int_{\mathcal{B}}d^{d}x\sqrt{-h}\left(K-\alpha\right)-\int_{\bm{\partial}\mathcal{B}}
d^{d-1}x \sqrt{-\Sigma}\, (\pi-\theta_0)\right)\Bigg].
\end{align}

In this formulation, the Newton constant in the bulk is represented by \( G_N^{(d+1)} \), while the AdS radius is denoted by \( L \). 
The terms \( K_{\bm{\partial}} \) and \( h_{\bm{\partial}} \) correspond to the extrinsic curvature and the induced metric of the conformal boundary \( \bm{\partial} \), respectively. 
Meanwhile, \( K \), \( h \), and \( \alpha \) denote the extrinsic curvature, the induced metric, and the tension component on the EOW brane \( \mathcal{B} \). 
On the intersection \( \partial\mathcal{B} \) between \( \mathcal{B} \) and \( \bm{\partial} \), the induced metric and the internal angle are expressed as \( \Sigma \) and \( \theta_0 \), respectively. 
Moreover, one can also introduce a Dvali-Gabadadze-Porrati (DGP) component \cite{Randall:1999vf,Dvali:2000hr} into this gravitational action to tune the effective Newton constant on the brane.

From a wider lens, this viewpoint is often termed as the ``bulk perspective''. There are also two alternative interpretations of this system from either the brane perspective or boundary perspective: 
\begin{description}
    \item [Brane perspective] The gravity of dimension \( (d+1) \) in the bulk finds its dual in a semi-classical gravity on the brane \( \mathcal{B} \), housing a CFT on both \( \mathcal{B} \) and the heat bath \( \bm{\partial} \) \cite{Gubser:1999vj}.
    \item [Boundary perspective] The combined gravity and matter field 
    on the brane can be mapped to a \( (d-1) \)-dimensional conformal defect system on \( \partial\mathcal{B} \) \cite{Almheiri:2019hni}. Within this frame of reference, the entire construction is encapsulated by the BCFT \cite{Takayanagi:2011zk,Chen:2020uac}.
\end{description}

The primary framework discussed in this section is expounded from the bulk perspective. 
To facilitate dynamical gravity on the brane, we set Neumann boundary conditions on the EOW brane (also refer to \cite{Miao:2017gyt,Chu:2017aab,Miao:2018qkc} for diverse perspectives). 
These conditions are articulated as:
\begin{equation}\label{eq:BCSonBrane_General}
  K_{ij}-K h_{ij}+\alpha h_{ij}=0,
\end{equation}
with \( h_{ij} \) denoting the induced metric on the brane \( \mathcal{B} \).


\subsection{The background}

However, the backreaction of the brane $\mathcal{B}$ to the bulk spacetime leads to a complicated bulk metric, which cannot be expressed analytically. For simplicity, in this paper, we consider the case with the vanishing of brane tension, i.e. $\alpha=0$. 

The AAdS spacetime with a brane is considered to be the TFD state at inverse temperature $\beta$, which leads to a compactification along imaginary time $\tau$, namely $\tau\sim\tau+\beta$.
From the brane perspective, this combined system is dual to an equilibrium system, where a two-sided black hole exchanges Hawking modes with heat baths. Here we consider a compactification along $y$ direction, namely $y\sim y+l$. Due to the compactification along $\tau$ and $y$ directions, we expect two solutions in bulk: the black hole solution and the thermal gas solution. As we will determine the solution below, the black hole solution becomes dominant when $\beta<l$, and the thermal gas solution becomes dominant when $\beta>l$. 

\begin{figure}
  \centering
  \subfigure[]{\label{fig:BHphase}
  \includegraphics[width=0.52\linewidth]{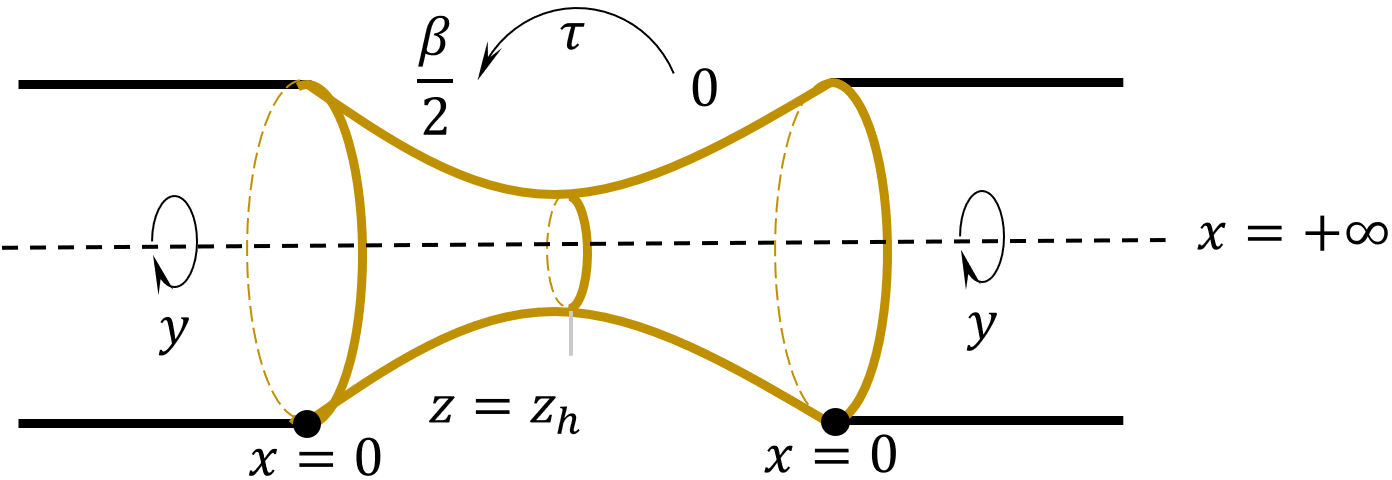}}
    \hspace{0pt}
  \subfigure[]{\label{fig:TGphase}
  \includegraphics[width=0.44\linewidth]{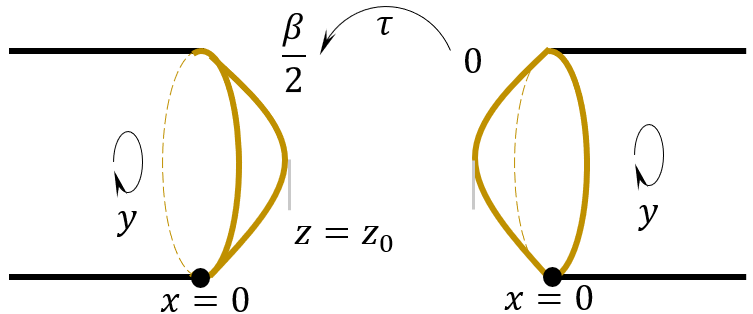}}
\caption{The spatial slices of the $(3+1)$-dimensional AAdS spacetime with the EOW brane depicted in brown. The ambient geometries show the (a) black hole solution and (b) the thermal gas solution.}\label{fig:GeometryPhase}
\end{figure}

The black hole solution renders a standard Schwarzschild-AdS metric as:
\begin{align}\label{eq:SAdS4_metric}
  ds^2=&\frac{1}{z^2}\left[-f(z)dt^2+\frac{dz^2}{f(z)}+dx^2+dy^2\right],\\
  f(z)=&1 -  \left(\frac{z}{z_h}\right)^3, \nonumber
\end{align}
where $\tau=it$, $0\leq y<l$ and $z_h=3\beta/4\pi$. To be consistent with the tension-vanishing condition, the positions of the conformal boundary and the EOW brane are respectively denoted as:
\begin{align}
\partial: \quad z=0 \qquad \text{and} \qquad \mathcal{B}:\quad x=0 \quad( \text{or equivalently}\;\theta_0=\pi/2).
\end{align}
The spatial slide of the black hole solution is shown in Fig.~\ref{fig:BHphase}.


It is noteworthy that \( z=z_h \) delineates the locus of the event horizon, which gives the Hawking temperature of the boundary combined system:
\begin{equation}
\frac1\beta=T_h=\frac{3}{4\pi z_h}.
\end{equation}
To determine the solution, we will calculate the on-shell action via the free energy. The renormalized action in $4$-dimensions can be expressed as \cite{PhysRevD.15.2752,Bianchi:2001kw,Elvang:2016tzz}
\begin{align}\label{eq:RenAction}
8\pi G_N^{(4)}I_{\text{ren}}=
& \frac{1}{2}\int d^{4}x
\sqrt{-g}\left(R+6\right)+\int_{\bm{\partial}}d^{3}x\sqrt{-h_{\bm{\partial}}}K_{\bm{\partial}}+\int_{\bm{\partial}}d^{3}x\sqrt{-h_{\bm{\partial}}}(R_{\bm{\partial}}+4)\nonumber\\
& + \textbf{brane terms}.
\end{align}
where we have set $L=1$ and $R_{\bm{\partial}}$ is the Ricci scalar associated with the induced metric $h_{\bm{\partial}}$ on the boundary. 
The dual Brown-York tensor for the boundary field theory can be obtained as
\begin{align}\label{eq:BYT}
\left\langle T_{ij}\right\rangle
=-2\lim_{z\rightarrow 0}\frac{1}{z\sqrt{h_{\bm{\partial}}}}\frac{\delta I_{\text{ren}}}{\delta (h_{\bm{\partial}})^{ij}}
=\frac1{8\pi G_N^{(4)}}\lim_{z\rightarrow0}\frac{1}{z}\left[G[h_{\bm{\partial}}]_{ij}-(K_{\bm{\partial}})_{ij}-\left(2-K_{\bm{\partial}}\right)(h_{\bm{\partial}})_{ij}\right],
\end{align}
with $G[h_{\bm{\partial}}]_{ij}$ being the Einstein tensor associated with $h_{\bm{\partial}}$.
With these expansions in hand, the free energy and the thermal entropy of the combined system can be expressed by
\begin{align}\label{eq:free energy}
	\mathcal{F}=E-T_h S_{h}, \quad S_{h}:=A_h/4G_N^{(4)}
\end{align}
with $E:=\int T_{00} dxdy$ the total energy of the combined system and $A_h$ the area of the event horizon. The black hole solution (\ref{eq:SAdS4_metric}) has the Euclidean on-shell action
\begin{align}\label{eq:free energy 1}
	I_{\text{BH}}=\beta\mathcal{F}_{\text{BH}}=-\frac{4\pi^2}{27 G_N^{(4)}} \frac{l}{\beta^2}L_x,
\end{align}
where $L_x$ is the regularized length in $x$ direction. 

The thermal gas solution can be obtained via the double Wick rotation of (\ref{eq:SAdS4_metric}) as
\begin{align}\label{eq:TG_metric}
  ds^2=&\frac{1}{z^2}\left[h(z)dy^2+\frac{dz^2}{h(z)}+dx^2-dt^2\right],\\
  h(z)=&1 -  \left(\frac{z}{z_0}\right)^3,\nonumber
\end{align}
with the periodicity $y\sim y+l$ and $\tau\sim \tau+\beta$ and without a horizon. Here $z_0=3l/4\pi$ and $S_h=0$. The spatial slide is shown in Fig.~\ref{fig:TGphase}. From the same holographic renormalization \eqref{eq:BYT} and \eqref{eq:free energy}, the thermal gas solution \eqref{eq:TG_metric} has the Euclidean on-shell action
\begin{align}\label{eq:free energy 2}
	I_{\text{TG}}=\beta\mathcal{F}_{\text{TG}}=-\frac{4\pi^2}{27 G_N^{(4)}} \frac{\beta}{l^2}L_x.
\end{align}

Comparing (\ref{eq:free energy 1}) with (\ref{eq:free energy 2}), we have
\begin{equation}
    I_{\text{BH}}-I_{\text{TG}}=
    -\frac{4\pi^2}{27 G_N^{(4)}}\kc{\frac{l}{\beta^2}-\frac{\beta}{l^2}}L_x.
\end{equation}
Therefore, for $T_h l > 1$, the system is in the black hole phase; while for $T_h l < 1$, the system is in the thermal gas phase. The entanglement proprieties are trivially static when the system is in the thermal gas phase. So we only focus on the entanglement properties when the system is in the black hole phase, and in the next subsection, we will evaluate the entanglement entropy of the radiation system first, in preparation for analyzing the mixed-state entanglement of the black hole interior.

\begin{figure}
  \centering
  \subfigure[]{\label{fig:TP}
  \includegraphics[width=0.48\linewidth]{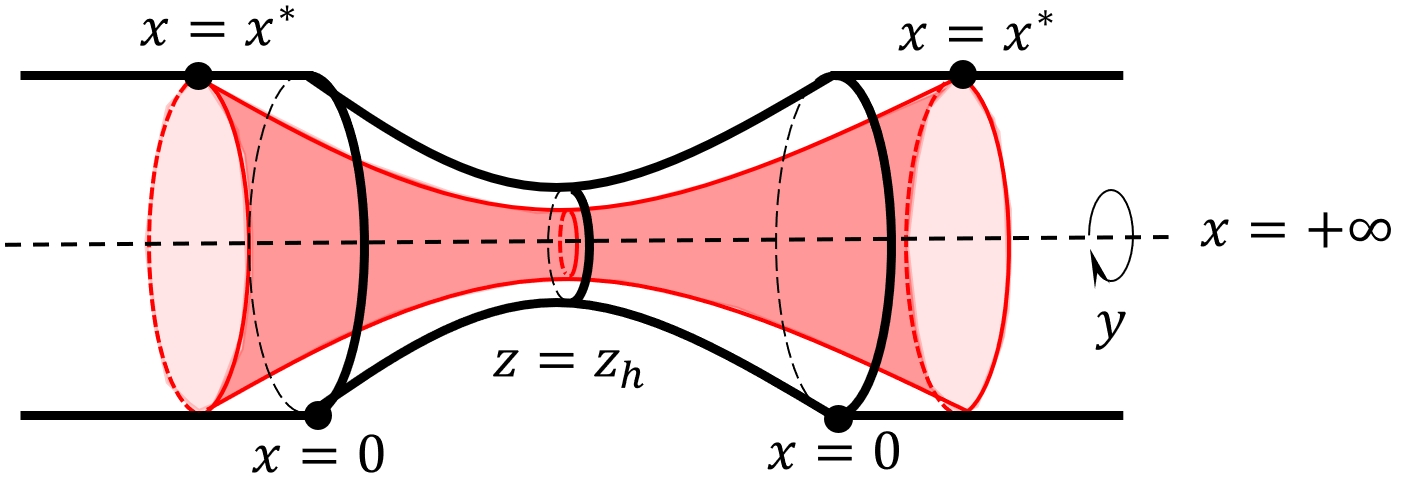}}
    \hspace{0pt}
  \subfigure[]{\label{fig:IP}
  \includegraphics[width=0.48\linewidth]{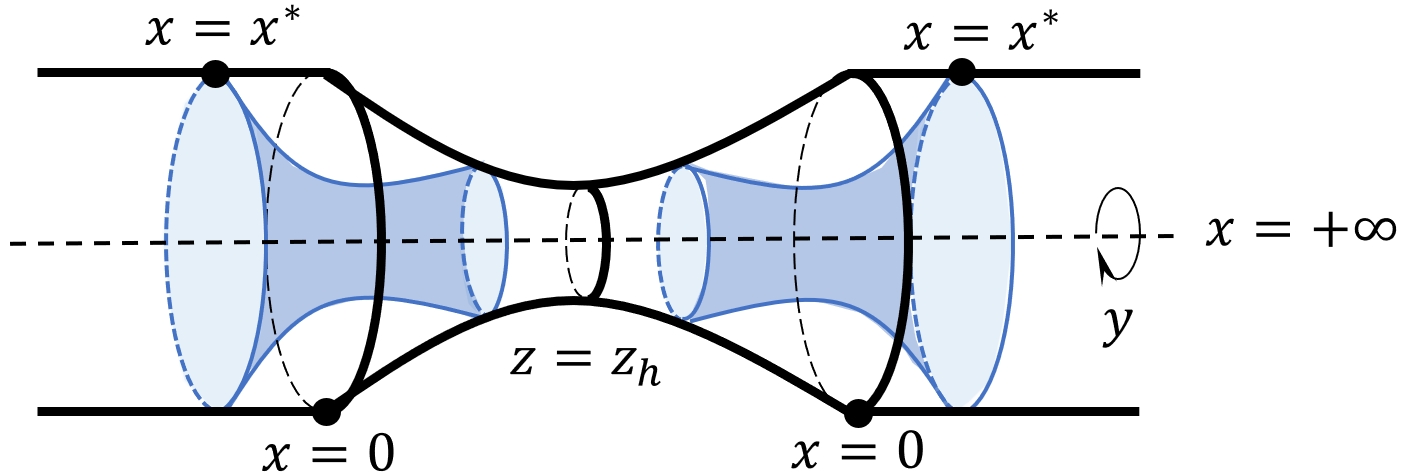}}
\caption{The HRT surfaces anchor at $x=x^*$ on the conformal boundary that separates the black hole system ($0\leq x \leq x^*$) and the radiation system ($x^* < x$). (a): The connected surface penetrates the event horizon. (b): The disconnected surface ends on the EOW brane.}\label{fig:T&IP}
\end{figure}
\subsection{The entanglement entropy of the radiation}
In this subsection, we will study the entanglement between the black hole and the radiation. First, we will construct the HRT surface of the radiation system to describe its entanglement entropy. Then, we will calculate the growth rate of the entanglement entropy.

From the brane perspective, we bipartition the whole system into black hole part $\mathcal{BH}$ and the radiation $\mathcal{R}$. Specifically, we refer to the region within $0\leq x\leq x^*$ as the black hole system $\mathcal{BH}$ \cite{Ling:2020laa}, while the remaining bath within $x>x^*$ as the radiation system $\mathcal{R}$, i.e.
\begin{align}\begin{split}
    \mathcal{BH}=&\left\{(z,x,y)|z=0,0\leq x \leq x^*,0 \leq y< l\right\}, \\
    \mathcal{R} =&\left\{(z,x,y)|z=0, x > x^*,0 \leq y< l\right\}.  
\end{split}\end{align}
For a fixed $l$, the larger $x^*$ is, the more degrees of freedom there are within the black hole. Intuitively, we will call a black hole containing more(less) degrees of freedom a larger(smaller) black hole.
\subsubsection{Plateau values}\label{sec:EE}
The entanglement between the black hole and radiation will be calculated by the HRT formula in a doubly holographic setup. In light of the Island rule, the entanglement entropy of the radiation system \( \mathcal{R} \) is determined by a quantum extremal surface. 
This can be dual to a \( (d+1) \)-dimensional HRT surface, given as \cite{Almheiri:2019psy,Chen:2020uac,Chen:2020hmv,Hernandez:2020nem}
\begin{align}\label{eq:SR}
S[\mathcal R]=\frac{1}{4G_N^{(d+1)}} \min_{\mathcal I} \left\{\mathop{\text{ext}}\limits_{\mathcal I} \left[\textbf{A}(\gamma_{\mathcal I\cup\mathcal R})\right]\right\},
\end{align}
where \( \gamma_{\mathcal I\cup\mathcal R} \) represents the HRT surface that shares its boundary with \( \mathcal I\cup\mathcal R \).
Within the framework of double holography, two distinct candidates emerge, each characterized by the HRT surface illustrated in Fig.~\ref{fig:T&IP}. 
One is the \textbf{connected surface} presenting an HRT surface \( \gamma_{T} \) anchored to both the left and right baths, and the island $\mathcal{I}$ is absent -- Fig.~\ref{fig:TP}; the other is the \textbf{disconnected surface} presenting a surface $\gamma_{I}$ anchored on the EOW brane $\mathcal{B}$ with a nontrivial island $\mathcal I$ on the brane. 
The presence of island \( \mathcal{I} \) ensures that the von Neumann entropy (\ref{eq:SR}) remains finite after the Page time \cite{Almheiri:2019yqk} -- Fig.~\ref{fig:IP}. Note that the HRT surface cannot pass through the $x=+\infty$ lines in Fig.~\ref{fig:T&IP}. Therefore, there are no other candidate HRT surfaces besides the two mentioned above.

On the one hand, when the entanglement entropy of the black hole system is described by a \textbf{disconnected surface}, it typically reaches a saturation point given by
\begin{equation}\label{eq:discon}
    S[\mathcal R]=\frac{1}{4G_N^{(4)}}\textbf{A}_{\text{dis}}:=\frac{2l}{4G_N^{(4)}} \int_\epsilon^{z_s} \frac{dz}{z^2}\sqrt{\frac{1}{f(z)}+y'(z)^2},
\end{equation}
with $y(z)$ representing the parameterization of the \textbf{disconnected surface} and $z_s$ being the turning point in $z$ direction. 
On the other hand, in cases where the entanglement entropy of the black hole system is represented by a \textbf{connected surface}, it increases over time, initially expressed at $t=0$ as
\begin{equation}\label{eq:con0}
    S[\mathcal R]\Bigg|_{t=0}=\frac{1}{4G_N^{(4)}}\textbf{A}_{\text{con}}:=\frac{2l}{4G_N^{(4)}} \int_\epsilon^{z_h} \frac{dz}{z^2}\sqrt{\frac{1}{f(z)}}.
\end{equation}
Based on these formulae, (\ref{eq:SR}) can be expressed as
\begin{equation}
    S[\mathcal R]=\frac{1}{4G_N^{(4)}} \min \left\{\textbf{A}_{\text{dis}}, \textbf{A}_{\text{con}}\right\}.
\end{equation}
\subsubsection{Evolution}
Investigating the dynamical evolution of entropy in doubly holographic models is typically challenging, mainly due to the complexity of solving the backreaction of the brane on the ambient geometry, which limits data accessibility inside the event horizon during evolution \cite{Ling:2020laa}. 
To circumvent these difficulties, in the current work, we adopt the probe limit approach, similar to \cite{Ling:2020laa}. 
In such scenarios, the EOW brane $\mathcal{B}$ does not exert backreaction on the ambient geometry, allowing the examination of time-dependent entanglement properties. 

When the entanglement entropy in a black hole system is characterized by the connected surface, it usually grows with time and can be written in the Eddington-Finkelstein coordinates $\{v,z,x,y\}$ as
\begin{equation}\label{eq:con0}
    S[\mathcal R]=\frac{l}{4G_N^{(4)}} \int \frac{ d\lambda}{z(\lambda)^{2}}\sqrt{-v'(\lambda ) \left[f(z(\lambda )) v'(\lambda )+2 z'(\lambda
  )\right]},
\end{equation}
with $\lambda$ being the intrinsic parameter of the connected surface (\ref{eq:con0}) and $$v=t-\int \frac{dz}{f(z)}.$$ 
Since the integrand in (\ref{eq:con0}) does not depend explicitly on $v$, one can derive a conserved quantity as
\begin{equation}\label{eq:C}
  C=\frac{f(z) v'+z'}{z^{2}\sqrt{-v' \left[f(z) v'+2 z'\right]}}.
\end{equation}
It is also noticed that the integral shown in (\ref{eq:con0}) is invariant under the reparametrization. 
Hence, the integrand can be chosen freely as
\begin{equation}\label{eq:C2}
  \sqrt{-v' \left[f(z) v'+2 z'\right]}=z^{2}.
\end{equation}  
Substituting (\ref{eq:C}) and (\ref{eq:C2}) into (\ref{eq:con0}), we have
\begin{align}
  \frac{d}{dt}S[\mathcal R]&=\frac{2l}{4G_N^{(4)}} \frac{\sqrt{-f(z_{max})}}{z_{max}^{2}},\\
  t&=\int_0^{z_{max}} dz \frac{C z^{2}}{f(z)\sqrt{f(z)+C^2z^{4}}}.
\end{align}
Here $z_{max}$ is the turning point of the connected surface and the relation between $z_{max}$ and the conserved quantity $C$ is given by
\begin{equation}
  f(z_{max})+C^2 z_{max}^{4}=0.
\end{equation}

At late times, the connected surface tends to surround a special extremal slice at $z=z_M$, as shown in
\cite{Hartman:2013qma}. 
Define
\begin{equation}
  F(z):=\frac{\sqrt{-f(z)}}{z^{2}},
\end{equation}
we find that $C^2=F(z_{max})^2$ keeps growing until meeting the extremum at $z_{max}=z_M$, where we have
  \begin{equation}\label{eq:Fp}
    F'(z_M) = -2 z_M^{-3}
    \sqrt{-f\left(z_M\right)}-\frac{z_M^{-2} f'\left(z_M\right)}{2\sqrt{-f\left(z_M\right)}} = 0.
  \end{equation}
By solving (\ref{eq:Fp}), we finally obtain the evolution of the entropy as
\begin{equation}\label{eq:rate}
  \lim_{t\rightarrow \infty} \frac{d}{dt} S[\mathcal R] = \frac{2l}{4G_N^{(4)}} F(z_M)= \frac{l}{4G_N^{(4)}} \frac{\sqrt{3}}{ \sqrt[3]{2}} T_h^{2}.
\end{equation}
For all the parameters considered in this paper, the $S[\mathcal R]$ grows from \eqref{eq:con0} with rate \eqref{eq:rate} and reach a plateau \eqref{eq:discon} at the Page time $\frac{\textbf{A}_{\rm dis}-\textbf{A}_{\rm con}}{4G}\Big/\left(\lim_{t\to\infty}\frac{dS[\mathcal R]}{dt}\right)$, as shown in Fig.~\ref{fig:evolution}.
In the following two subsections, we will study the entanglement between the bi-partition of the black hole, by calculating the holographic reflected entropy and mutual information respectively.
\subsection{The reflected entropy of the bipartite black hole}
To study the entanglement inside the black hole, We further partition the black hole into two subsystems as
\begin{align}\begin{split}
    \mathcal{BH}_1=&\left\{(z,x,y)|z=0,0\leq x \leq x^*,0\leq y < l/2 \right\},  \\
    \mathcal{BH}_2=&\left\{(z,x,y)|z=0,0\leq x \leq x^*,l/2 \leq y < l \right\}. 
\end{split}\end{align}
\subsubsection{Plateau values}
To quantify this mixed-state entanglement, we turn to calculate the holographic reflected entropy \cite{Dutta:2019gen,Engelhardt:2022qts}, which is proposed to be measured by the area of the quantum entanglement wedge cross section (Q-EWCS). 

Beyond the quantum extremal surface, we can introduce the Q-EWCS to delineate the contributions of quantum fields within the entanglement wedge.
This can be further dual to the classical EWCS in a one-dimensional higher spacetime in the doubly holographic setup \cite{Ling:2021vxe,Liu:2023ggg}. 
The EWCS between $\mathcal{BH}_1$ and $\mathcal{BH}_2$ is defined as the minimal codimension-two cross-section in the entanglement wedge, which can be articulated as:
\begin{equation}
S^R[\mathcal{BH}_1:\mathcal{BH}_2] = \frac{2}{4G_N^{(d+1)}}\textbf{A}(E[\mathcal{BH}_1:\mathcal{BH}_2]),
\end{equation}
where \( E[\mathcal{BH}_1:\mathcal{BH}_2] \) stands for the EWCS dividing the entanglement wedge of \(\mathcal{BH}_1\) and \(\mathcal{BH}_2\). Notably, it is characterized by having a minimal area and is bounded by the HRT surface and EOW brane. 

\begin{figure}
  \centering
  \subfigure[]{\label{fig:T1}
  \includegraphics[width=0.48\linewidth]{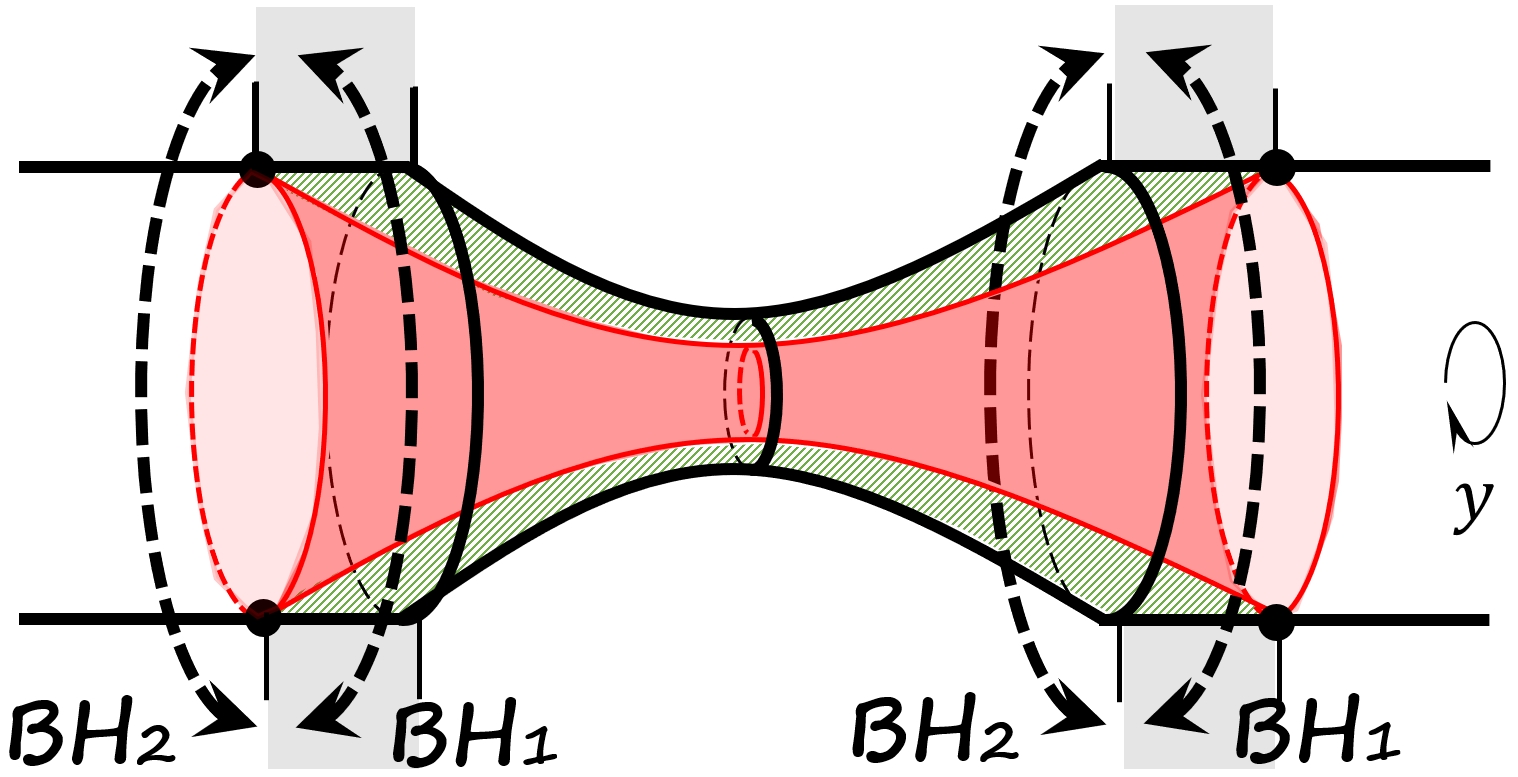}}
    \hspace{0pt}
  \subfigure[]{\label{fig:T2}
  \includegraphics[width=0.48\linewidth]{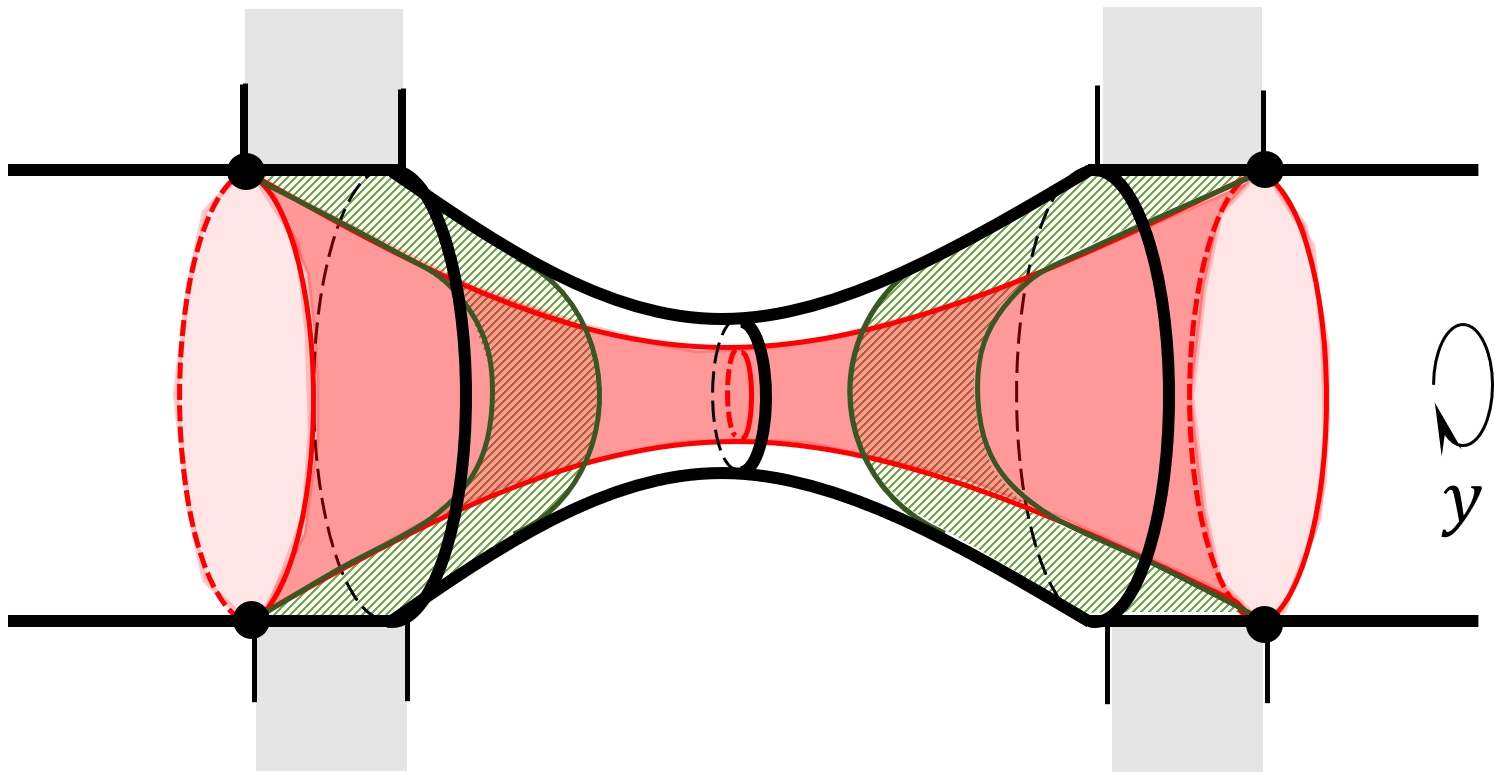}}
    \hspace{0pt}
  \subfigure[]{\label{fig:I1}
  \includegraphics[width=0.48\linewidth]{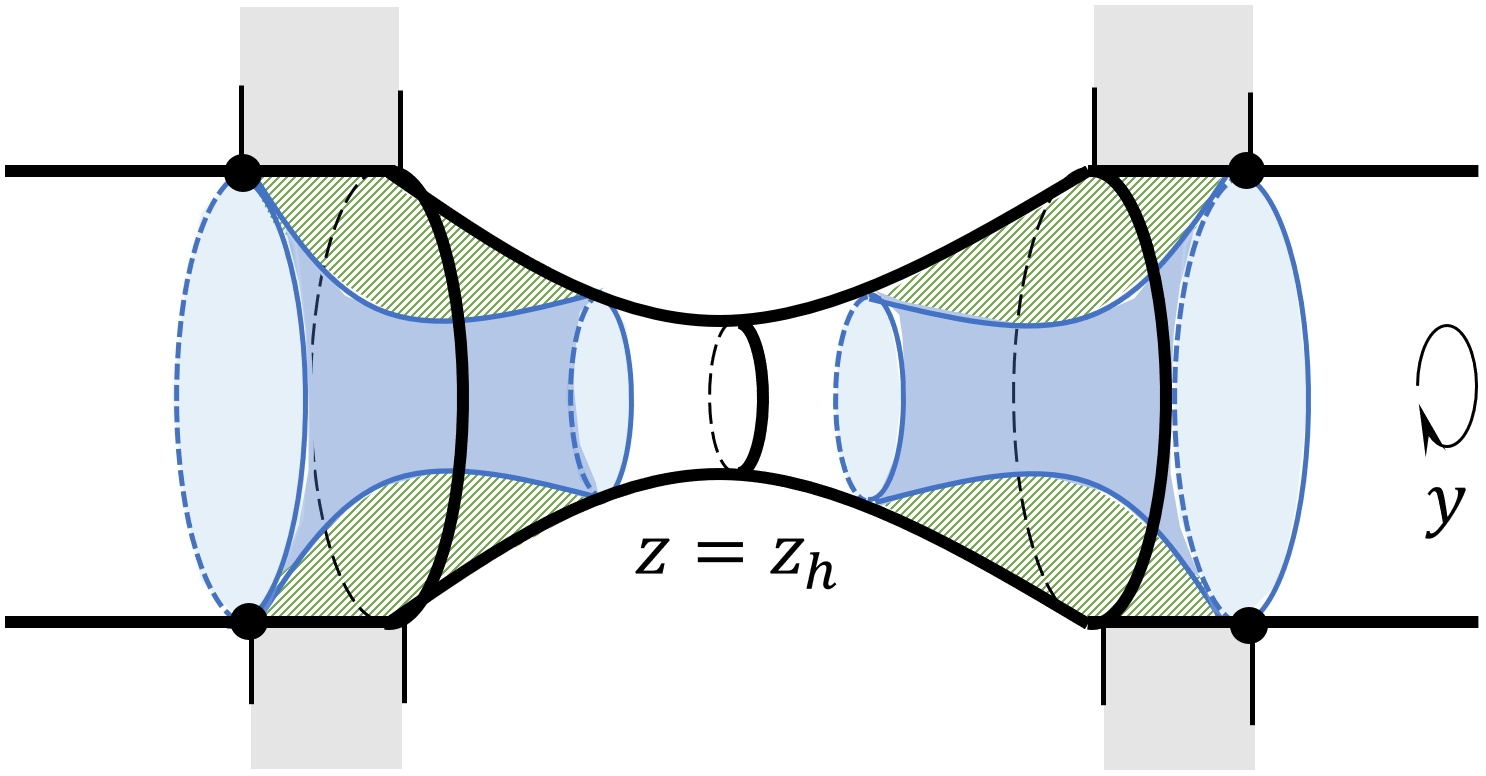}}
\caption{The EWCS represented by the 2-dimensional green surface between the black hole bipartition \(\mathcal{BH}_1\) and \(\mathcal{BH}_2\) in (a) the first trivial phase Ph-\textbf{T}\(_1\), (b) the second trivial phase Ph-\textbf{T}\(_2\), and (c) the island phase Ph-\textbf{I}.}\label{fig:RE_T&IP}
\end{figure}

When the HRT surface of the black hole system is dominated by the \textbf{connected surface}, there exist two possible phases for \( E[\mathcal{BH}_1:\mathcal{BH}_2] \). 
The first phase we call Ph-\textbf{T}\(_1\), is a co-dimension two surface, which pierces the event horizon and ends on the opposite conformal boundary as depicted in Fig.~\ref{fig:T1}. As this surface evolves temporally, the correlation between \(\mathcal{BH}_1\) and \(\mathcal{BH}_2\) correspondingly increases, which can be initially given at $t=0$ by
\begin{equation}
S^R[\mathcal{BH}_1:\mathcal{BH}_2]\Bigg|_{t=0} = \frac{8x^*}{4G_N^{(4)}}\int_\epsilon^{z_h} \frac{dz}{z^2}\sqrt{\frac{1}{f(z)}}.
\end{equation}
Alternatively, the second phase that we call Ph-\textbf{T}\(_2\), is a co-dimension-two surface that terminates on the same side of the conformal boundary, as illustrated in Fig.~\ref{fig:T2}. 
The corresponding reflected entropy can be expressed as
\begin{equation}
S^R[\mathcal{BH}_1:\mathcal{BH}_2] = \frac{8x^*}{4G_N^{(4)}} \int_\epsilon^{z_s} \frac{dz}{z^2}\sqrt{\frac{1}{f(z)}+y'(z)^2},
\end{equation}
where $y(z)$ represents the parameterization of the \text{Ph}-\textbf{T}$_2$, and $z_s$ is the turning point in $z$ direction. 
The green minimal cross-section in Fig.~\ref{fig:T2}, representing a saddle point, spontaneously breaks the $Z_2$ symmetry between the subregions $\mathcal{BH}_1$ and $\mathcal{BH}_2$. This green surface shifts outward or backward relative to the red surface, towards the viewer's perspective. The degeneracy can be lifted by slightly breaking the symmetry between the sizes of the subregions $\mathcal{BH}_1$ and $\mathcal{BH}_2$. A similar phenomenon occurs in the RT formula for a disjoint entanglement region, where the connected RT surface and the disconnected RT surface have equivalent areas.

When the HRT surface of the black hole system is described by the disconnected surface, we turn to the island phase (Ph-\textbf{I}), where the \( E[\mathcal{BH}_1:\mathcal{BH}_2] \) remains static, reflecting the invariance of the correlation between $\mathcal{BH}_1$ and $\mathcal{BH}_2$ -- Fig.~\ref{fig:I1}. 
The corresponding reflected entropy is
\begin{equation}
S^R[\mathcal{BH}_1:\mathcal{BH}_2] = \frac{8}{4G_N^{(4)}} \int_\epsilon^{z_s} \frac{dz}{z^2}\sqrt{\frac{1}{f(z)}}x(z).
\end{equation}

\subsubsection{Evolution}
When \( \mathcal{BH}_1\) and \(\mathcal{BH}_2 \)  are in Ph-\textbf{T}\(_1\), the reflected entropy can be described by the EWCS \( E[\mathcal{BH}_1:\mathcal{BH}_2] \) penetrating the horizon also in the Eddington-Finkelstein coordinates as
\begin{equation}\label{eq:phaseT1}
    S^R[\mathcal{BH}_1:\mathcal{BH}_2]=\frac{4x^*}{4G_N^{(4)}} \int \frac{ d\lambda}{z(\lambda)^{2}}\sqrt{-v'(\lambda ) \left[f(z(\lambda )) v'(\lambda )+2 z'(\lambda
  )\right]},
\end{equation}
with the Lagrangian being calculated in the Eddington-Finkelstein coordinates and the growth rate during the evolution can be similarly measured by
\begin{align}
  \frac{d}{dt}S^R[\mathcal{BH}_1:\mathcal{BH}_2]&=\frac{8x^*}{4G_N^{(4)}} \frac{\sqrt{-f(z_{max})}}{z_{max}^{2}},\\
  t&=\int_0^{z_{max}} dz \frac{C z^{2}}{f(z)\sqrt{f(z)+C^2z^{4}}}.\nonumber
\end{align}

Then, we will calculate the dynamical evolution of the reflected entropy.
For numerical convenience, we set the UV cutoff near the conformal boundary at $z=\epsilon=1/100$. 
The analysis utilizes dimensionless parameters $\{y^*/x^*, t/x^*, T_h x^*,\epsilon/x^*\}$,where $y^*=l/4$. Specifically, to describe the properties of different evolution patterns, we subtract the initial value of the entanglement measures as 
\begin{equation}
    \Delta \mathcal{M}:=\mathcal{M} -\mathcal{M}|_{t=0},
\end{equation}
where $\mathcal{M}\in \{S[\mathcal{R}],S[\mathcal{BH}_1],S[\mathcal{BH}_2],S^R[\mathcal{BH}_1:\mathcal{BH}_2],I[\mathcal{BH}_1:\mathcal{BH}_2]\}$,
 with $S[\mathcal{BH}_1]$, $S[\mathcal{BH}_2]$ and $I[\mathcal{BH}_1:\mathcal{BH}_2]$ denoting the entanglement entropy of $\mathcal{BH}_1$, $\mathcal{BH}_2$ and the mutual information of the black hole bipartition.

\begin{figure}
  \centering
  \subfigure[]{\label{fig:EEini}
  \includegraphics[width=0.47\linewidth]{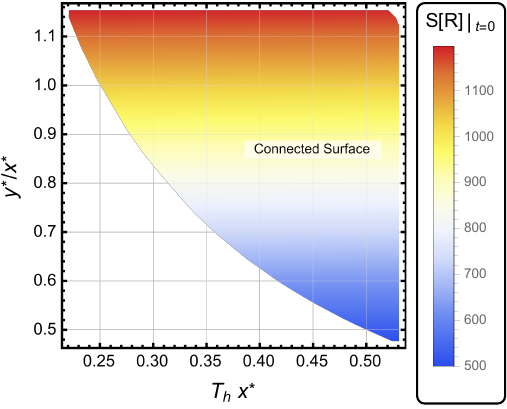}}
    \hspace{0pt}
  \subfigure[]{\label{fig:REini}
  \includegraphics[width=0.49\linewidth]{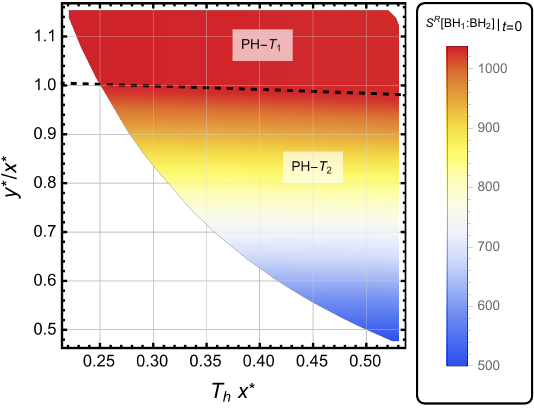}}
\caption{(a): The entanglement entropy of the radiation system at $t=0$. (b): The reflected entropy of the black hole bipartition at $t=0$. In addition, we have set $4G_N^{(4)}=1$. The blank region of $T_h 4y^*<1$ corresponds to the thermal gas phase, which is excluded.}\label{fig:EE&RE}
\end{figure}

Firstly, we observe that the initial phase diagram of the reflected entropy of the black hole bipartition largely depends on the configuration of the HRT surface of the corresponding black hole system $\mathcal{BH}$. 
Initially, the entanglement entropy of $\mathcal{BH}$ is shown in Fig.~\ref{fig:EEini}. 
In general, hotter black holes are more likely to have their entanglement entropy with the radiation represented by a connected surface that grows over time. 
In contrast, for cooler black holes, the entanglement entropy typically remains static, described by a disconnected surface, which only dominates at $t=0$ for $T_h x^*<0.14$ and is not covered by the range of parameters in this figure.
Importantly, the transition from connected to disconnected surfaces will also trigger a phase transition for the reflected entropy, as the EWCS is confined by the RT surface.

Specifically, for a hotter and larger black hole, the EWCS tends to stay in the $T_1$ phase (Ph-\textbf{T}\(_1\)), while for a cooler and smaller black hole, the EWCS tends to stay in the $T_2$ phase (Ph-\textbf{T}\(_2\)). The island phase only dominates at $t=0$ for $T_h x^*<0.14$, and hence is not covered.
The reason is obvious from the viewpoint of gravity. 
For a hotter black hole, the event horizon is closer to the conformal boundary. 
Its influence on the EWCS is twofold: the area of the EWCS in Ph-\textbf{T}\(_1\) is compressed by the location of the horizon, while the area of the EWCS in Ph-\textbf{T}\(_2\) becomes larger due to the volume law entanglement.  

\begin{figure}
  \centering
  \subfigure[]{\label{fig:PhaseDiagram}
  \includegraphics[width=0.32\linewidth]{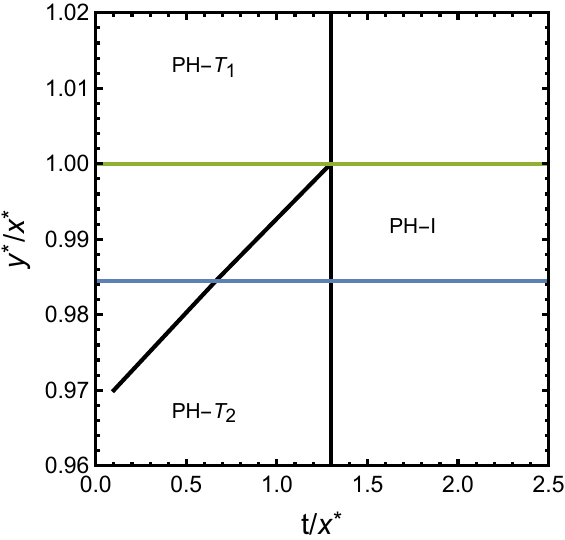}}
    \hspace{0pt}
  \subfigure[]{\label{fig:Type2}
  \includegraphics[width=0.31\linewidth]{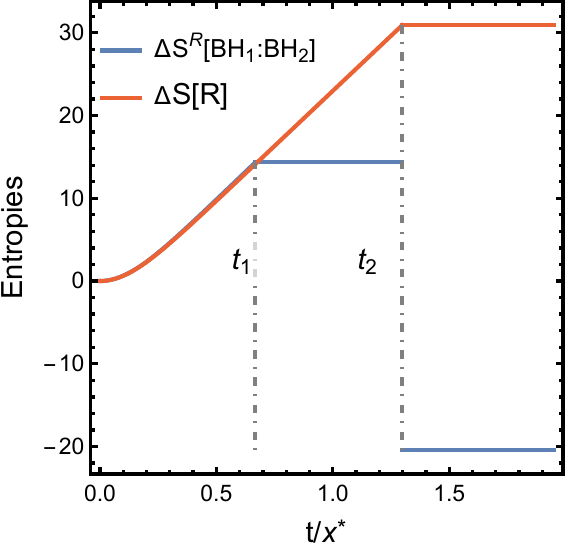}}
    \hspace{0pt}
  \subfigure[]{\label{fig:Type3}
  \includegraphics[width=0.31\linewidth]{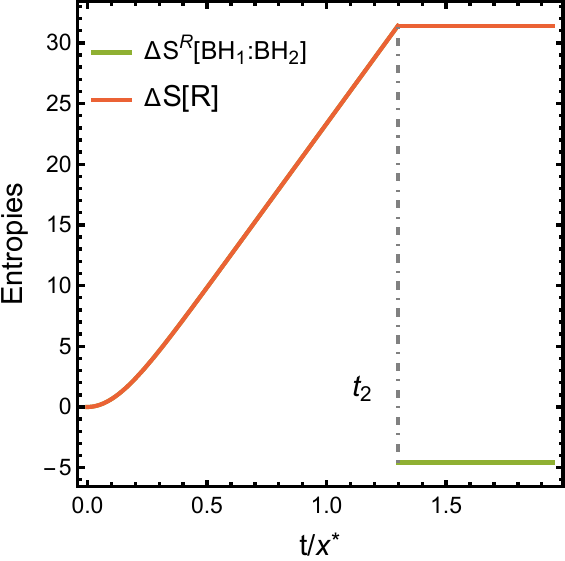}}
\caption{The different evolutionary patterns of the entropies with the dimensionless temperature $T_hx^* = 0.530$. (a) Ph-\textbf{T}$_1$, Ph-\textbf{T}$_2$ and Ph-\textbf{I} of the reflected entropy of the black hole bipartition are bounded by the black curves. We plot the evolution of entropies along the green and blue trajectories in other panels. (b) For small black holes with $y^*/x^* = 0.985$, the reflected entropy exhibits a ramp-plateau-slump behavior. (c): For large black holes with $y^*/x^* = 1.000$, the reflected entropy exhibits a ramp-slump behavior. The slumps appear at the Page time $t_2$. The red curves represent the entanglement entropy of the radiation system.}\label{fig:evolution}
\end{figure}

Next, we observe that the dynamical evolution and the possible transition of the reflected entropy of the black hole bipartition depend on the dynamical behavior of the entanglement entropy as well. 
We show that the phase transition of the EWCS occurs during the evolution when the RT surface is initially the connected surface -- Fig.~\ref{fig:REini}. 
Fig.~\ref{fig:REini} illustrates the entanglement phase diagram of the EWCS at the beginning of evolution $t=0$. 
Subsequently, this diagram will generally evolve over time. 
Specifying the dimensionless temperature as $T_hy^*= 0.53$, the corresponding dynamical phase structure during the evolution is shown in Fig.~\ref{fig:evolution}. 
Different phases are undergone by the reflected entropy of the black hole bipartition during the dynamical evolution -- Fig.~\ref{fig:PhaseDiagram}, which depend on different ratios $y^*/x^*$ of the bipartition. The black curves separate the diagram into Ph-\textbf{T}$_1$, Ph-\textbf{T}$_2$ and Ph-\textbf{I} phases.

By virtue of the subtracted entanglement entropy of the radiation system and the reflected entropy of the black hole bipartition, we may extract two typical evolution patterns as follows.

\begin{description}
\item[Type I: the evolution with plateau.]
  In a smaller black hole, the reflected entropy $S^R[\mathcal{BH}_1:\mathcal{BH}_2]$ typically exhibits a ramp-plateau-slump behavior with two phase transitions at $t_1$ and $t_2$, as shown in Fig.~\ref{fig:Type2}.
  When $t/x^*<t_1$, the reflected entropy in Ph-\textbf{T}$_1$ intensifies owing to the internal interactions of the black hole system. 
  When $t_1<t/x^*<t_2$, the reflected entropy in Ph-\textbf{T}$_2$ achieves saturation. When $t_2<t/x^*$, the entanglement entropy for the entire black hole is given by the disconnected RT surface, and hence the entanglement of the entire system achieves equilibrium. This leads to a discontinuous slump of the reflected entropy in Ph-\textbf{I}.

  \item[Type II: the evolution without plateau.]
  In a larger black hole, the reflected entropy $S^R[\mathcal{BH}_1:\mathcal{BH}_2]$ typically exhibits a ramp-slump behavior with a single phase transition at $t_2$, as depicted in Fig.~\ref{fig:Type3}.
  The reflected entropy directly transitions from Ph-\textbf{T}$_1$ to Ph-\textbf{I} without having a plateau.
\end{description}

For both types of evolutionary behavior, we observe that the final value of the reflected entropy after the black hole Page time $t_2$ is always smaller than its initial value, indicating the strong thermalization of the black hole with the radiation, regardless of the amount of entanglement generated by the internal interaction.

At the end of this subsection, we briefly comment on the effect of the radiation $\mathcal R$ on the entanglement inside the black hole $\mathcal{BH}$. To explore the dynamics in isolation from the radiation, one can conceptually decouple the black hole from the radiation by placing another EOW brane at $x=x^*$. We will consider a large enough $x^*$ such that the EOW branes located at $x=0$ and $x=x^*$ are disconnected in order to allow the entanglement to grow. Intriguingly, after this decoupling from the radiation, the entanglement wedge of the black hole $\mathcal{BH}$ remains the same as observed in the coupled scenarios, except that the island phase as well as the Page time cease to exist. Consequently, the reflected entropy between the bipartition $\mathcal{BH}_1:\mathcal{BH}_2$ grows as fast as it does in the coupled scenarios.
This phenomenon suggests that the early thermalization driven by the radiation does not significantly influence the entanglement inside the black hole.

\subsection{The mutual information of the bipartite black hole}
Another mixed-state entanglement measure, mutual information, is also extensively discussed in the literature. For any two individual systems $\mathcal{BH}_1$ and $\mathcal{BH}_2$, their mutual information can be expressed as
\begin{equation}\label{eq_MI}
  I[\mathcal{BH}_1:\mathcal{BH}_2]=S[\mathcal{BH}_1]+S[\mathcal{BH}_2]-S[\mathcal{R}].
\end{equation}
Here, the explicit expression of the last term $S[\mathcal{R}]$ is just the entanglement entropies shown in Section \ref{sec:EE}, and since we always focus on the cases with $S[\mathcal{BH}_1]=S[\mathcal{BH}_2]$, we will only derive the expression of $S[\mathcal{BH}_1]$. In this subsection, we analyze the mutual information inside the black hole during evolution.

  \begin{figure}
  \centering
  \subfigure[]{\label{fig:trivialsub}
  \includegraphics[width=0.31\linewidth]{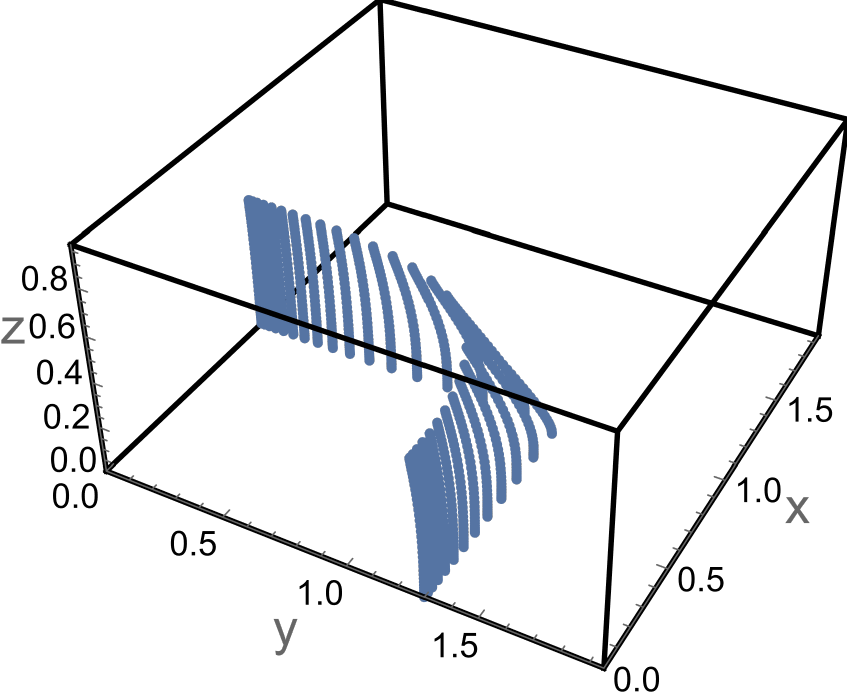}}
    \hspace{0pt}
  \subfigure[]{\label{fig:islandsub}
  \includegraphics[width=0.31\linewidth]{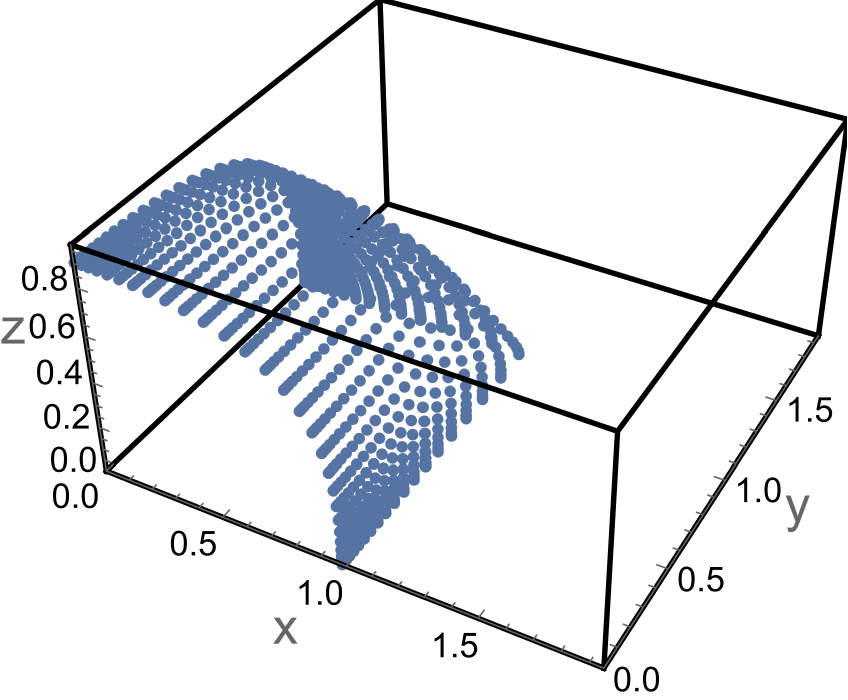}}
\caption{The discrete version of the candidates of the (half of) HRT surface of the black hole subsystem $\mathcal{BH}_1$ ($\mathcal{BH}_2$). (a) The first candidate $\gamma_T$ is illustrated with parameters $\{x^*,y^*,z_h\}=\{64/65,13/10,1/5^{1/3}\}$.  (b) The second candidate $\gamma_I$ is illustrated with parameters $\{x^*,y^*,z_h\}=\{64/65,13/10,1\}$. Here $y$ direction is periodic with $y\sim y+l$, with $l=4y^*$.}\label{fig:subphase}
\end{figure}

\subsubsection{Plateau values}
In general, there are two candidates for the HRT surface of $\mathcal{BH}_1$. The first candidate is a co-dimension two surface $\gamma_T$ that goes across the event horizon -- Fig.~\ref{fig:trivialsub}, and hence grows with time. At the initial stage $t=0$, the area functional of $\gamma_T$ can be expressed as
\begin{equation}\label{eq:t}
    \textbf{A}(\gamma_T)=4\int_{\epsilon}^{z_h} dz \int_0^{\pi/2} d\theta \frac{1}{z^2}\sqrt{\frac{r(\theta ,z)^2 \left[f(z) r^{(0,1)}(\theta ,z)^2+1\right]+r^{(1,0)}(\theta ,z)^2}{f(z)}}.
\end{equation}
Here $\{r,\theta,z\}$ is a three-dimensional cylindrical-like coordinate set to describe the surface $\gamma_{T}$, with $r$ being the radial distance in $z=const.$ plane, $\theta$ being the azimuthal angle  in $z=const.$ plane, measured from the positive $y$-axis, by the coordinate transformation $$x=r \sin \theta,\qquad y=r \cos\theta.$$
The corresponding boundary conditions of $r(\theta,z)$ are given by Tab.~\ref{tab:bcs_t1}.
\begin{table}
  \centering
  \begin{tabular}{|c| c| c |c |}
    \hline
      $\mathbf{\theta=0}$ & $\mathbf{\theta=\pi/2}$ & $\mathbf{z=0}$ & $\mathbf{z=z_h}$  \\
    \hline
    $\partial_{\theta} r =0$ & $\partial_{\theta} r =0$ & $1/r=\sqrt[n]{\left(\frac{\cos\theta}{x^*}\right)^n +\left(\frac{\sin\theta}{y^*}\right)^n}$ & $\partial_{z} r =0$ \\
    \hline
  \end{tabular}
  \caption{The boundary conditions. For sufficient large $n$, the configuration of $r(\theta,0)$ on the conformal boundary approaches $\mathcal{BH}_i$, where $i=1,2$. For numerical simulation, we choose $n=100$.}\label{tab:bcs_t1}
\end{table}

The second candidate $\gamma_I$ dominates after the Page time of $\mathcal{BH}_1$, which remains still -- Fig.~\ref{fig:islandsub}. Here we can use three-dimensional spherical-like coordinates $\{ r',\theta',\phi \}$ to describe this surface $r'(\theta',\phi)$, with the coordinate transformation
$$z=r'\sin\theta'\cos\phi,\qquad x=-r' \cos \theta'\cos\phi,\qquad y=r' \sin\phi.$$
The area functional of $\gamma_I$ can be expressed as
\begin{equation}\label{eq:i}
    \textbf{A}(\gamma_I)=4\int_{0}^{\pi/2} d\phi \int_{\theta_\epsilon}^{\pi/2} d\theta' \mathcal{L}[r'(\theta',\phi),\partial_{\theta'}r',\partial_{\phi}r',\theta',\phi],
\end{equation}
where $\theta_\epsilon$ is obtained by solving
$$z_\epsilon=r'(\theta'_\epsilon,\phi)\sin\theta'_\epsilon\cos\phi$$
at any specific $\phi$, and the explicit expression of $\mathcal{L}$ is so long that we do not write it.
The corresponding boundary conditions of $r'(\theta',\phi)$ are given by Tab.~\ref{tab:bcs_i}.
\begin{table}
  \centering
  \begin{tabular}{|c| c| c |c |}
    \hline
      $\mathbf{\theta'=\pi/2}$ & $\mathbf{\theta'=\pi}$ & $\mathbf{\phi=0}$ & $\mathbf{\phi=\pi/2}$  \\
    \hline
    $\partial_{\theta'} r' =0$  & $1/r'=\sqrt[n]{\left(\frac{\cos\phi}{x^*}\right)^n +\left(\frac{\sin\phi}{y^*}\right)^n}$  & $\partial_{\phi} r' =0$ & $r' =y^*$ \\
    \hline
  \end{tabular}
  \caption{The boundary conditions. For sufficient large $n$, the configuration of $r'(\pi/2,\phi)$ approaches $\mathcal{BH}_i$, where $i=1,2$. Here we also choose $n=100$ for numerical simulations.}\label{tab:bcs_i}
\end{table}

\subsubsection{Evolution}
When the entanglement entropies of \( \mathcal{BH}_1\) and \(\mathcal{BH}_2 \)  are described by $\gamma_T$, its growth can be approximated by
\begin{align}\label{eq:growth_mi}
  \frac{d}{dt}S[\mathcal{BH}_1]&\approx\frac{2(2x^*+y^*)}{4G_N^{(4)}} \frac{\sqrt{-f(z_{max})}}{z_{max}^{2}},\\
  t&=\int_0^{z_{max}} dz \frac{C z^{2}}{f(z)\sqrt{f(z)+C^2z^{4}}}.\nonumber
\end{align}
Note that due to the constraints of our numerical strategy, we provide only an approximation of the growth rate. An exact solution may be derived within the framework of Eddington-Finkelstein coordinates, which would facilitate the description of a series of two-dimensional inhomogeneous extremal surfaces.

  \begin{figure}
  \centering
  \includegraphics[width=0.55\linewidth]{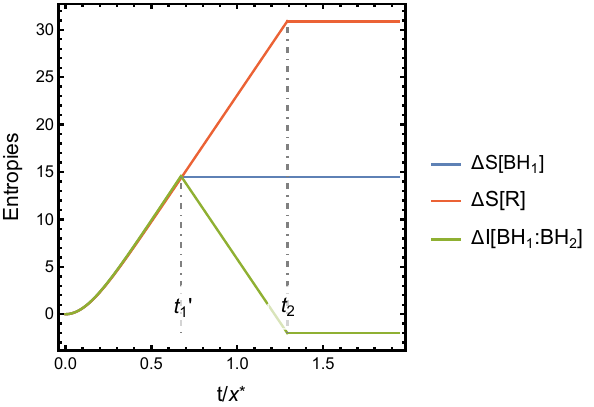}
\caption{The growth of different entanglement measures. The red curve represents the entanglement entropy of the radiation system, while the blue curve represents the entanglement entropy of the black hole subsystem, and the green curve stands for the mutual information between the bipartition of the black hole system. $t'_1$ is the Page time of the black hole subsystem, while $t_2$ is the Page time of the black hole system, with the dimensionless parameters to be $\{T_h x^*, y^*/x^*\}=\{0.530, 0.985\}$.}\label{fig:mi}
\end{figure}

The mutual information exhibits a ramp-slope-stabilizing behavior, as shown in Fig.~\ref{fig:mi}. There are two distinct times scales $t'_1$ and $t_2$, standing for the Page time of the black hole bipartition and the entire black hole system, respectively. 
The ramp is generated by the internal interaction. The slope is due to the saturation of the entanglement entropy of the subsystem $\mathcal{BH}_{1,2}$. The stabilizing phase indicates the final thermalization of the black hole with the radiation, where the black hole subsystem is primarily entangled with the radiation rather than with the rest of the black hole. After all, although our approximation on the growth rate may affect the accuracy of the value of $t_1'$, the analysis of the evolution of the mutual information remains qualitatively reliable.

Let us compare the reflected entropy and the mutual information of the bipartition \( \mathcal{BH}_1:\mathcal{BH}_2 \). First, under our approximation, their initial growth rates are the same, both being proportional to the area of the interface of the bipartition. Second, the first transition time \( t_1 \) of the reflected entropy and \( t'_1 \) of the mutual information are close but not identical in principle, since these two timescales are determined by comparing different sets of minimal surface candidates. Third, the reflected entropy grows or reaches a plateau, while the mutual information decreases just before the black hole Page time \( t_2 \). Fourth, both of their final values are smaller than their initial values, respectively. Both interior entropy quantities describe the same tendency: entanglement is initially created within the system through internal interactions and eventually dissipates into the surrounding environment.


\section{The SYK cluster coupled to a bath}\label{sec:SYK}

In this section we turn to study the evolution of entanglement inside the SYK cluster coupled to a bath. 
The approaches we consider here are generalized from \cite{Zhang:2019evaporation,Gu:2017njx,Gu:2016oyy,Penington:2019kki,Almheiri:2019jqq,Su:2020quk}. We expect the interior entanglement will grow initially and decay to a low value after the Page time of the whole SYK cluster, similar to the tendency in gravity. 

\subsection{Hamiltonian}

We consider a SYK cluster $\psi_j$ with $j=1,2,\cdots,N$ coupled to a huge SYK bath $\phi_{j'}$ with $j'=1,2,\cdots,M$. The Majorana fermions $\ke{\psi_j,\phi_{j'}}$ obey the anti-commutation relation $\ke{\psi_j,\psi_k}=\ke{\phi_j,\phi_k}=\delta_{jk}$. 
The Hamiltonian of the whole system is
\begin{align}
    H=i^{q/2}\sum_{I_q}^N J_{I_q} \Psi_{I_q} + i^{p/2}\sum_{I_p'}^M K_{I_p'} \Phi_{I_p'} + i^{(n+m)/2}\sum_{I_s}^N\sum_{I_r'}^M V_{I_sI_r'}\Psi_{I_s}\Phi_{I_r'}, \label{eq:SYKHam}
\end{align}
where we use the shorthand $I_q=j_1j_2\cdots j_q$, $I_p'=j_1'j_2'\cdots j_p'$, $\Psi_{I_q}=\psi_{j_1}\psi_{j_2}\cdots \psi_{j_q}$, $\Phi_{I_p'}=\phi_{j_1'}\phi_{j_2'}\cdots \phi_{j_p'}, \sum_{I_q}^N=\sum_{1\leq j_1< j_2<\cdots<j_q\leq N}$, and numbers $q,p,s+r$ are positive and even. The disorder couplings $\ke{J_{I_q},K_{I'_q},V_{I_sI'_r}}$ have zero mean values and finite variances 
\begin{align}
    \avg{J_{I_q}^2}=2^{q-1}\frac{(q-1)!}{N^{q-1}}J^2,\quad 
    \avg{K_{I_p'}^2}=2^{p-1}\frac{(p-1)!}{N^{p-1}}K^2,\quad 
    \avg{V_{I_sI_r'}^2}=2^{s+r-1}\frac{(s-1)!r!}{N^{s-1}M^r}V^2.
\end{align}
We will work on the effective action with bi-local fields. We identify
\begin{align}
    G_\psi(\tau_1,\tau_2)=\frac1{N}\sum_{j=1}^{N} \psi_j(\tau_1)\psi_j(\tau_2),\quad 
    G_\phi(\tau_1,\tau_2)=\frac1{M}\sum_{j=1}^{M} \phi_j(\tau_1)\phi_j(\tau_2),
\end{align}
by introducing the Legendre multipliers $\ke{\Sigma_\psi(\tau_1,\tau_2),\, \Sigma_\phi(\tau_1,\tau_2)}$ in the path integral.
Integrating out the disorder and Majorana fermions, we obtain the effective action for the bi-local fields in the replica diagonal part
\begin{align}\label{SYKAction}
    \S=&~\S_\psi+\S_\phi+\S_i,\\ 
    -\S_\psi=&~N\frac12\log\det(\partial-\Sigma_\psi)+N\int_0^\beta d\tau_1d\tau_2 \kd{-\frac12\Sigma_\psi(\tau_1,\tau_2) G_\psi(\tau_1,\tau_2)+\frac{J^2}{4q}(2G_\psi(\tau_1,\tau_2))^q}, \nn\\
    -\S_\phi=&~M\frac12\log\det(\partial-\Sigma_\phi)+M\int_0^\beta d\tau_1d\tau_2 \kd{-\frac12\Sigma_\phi(\tau_1,\tau_2) G_\phi(\tau_1,\tau_2)+\frac{K^2}{4p}
(2G_\phi(\tau_1,\tau_2))^p}, \nn\\
    -\S_i=&~N\frac{V^2}{4s}\int_0^\beta d\tau_1d\tau_2 (2G_\psi(\tau_1,\tau_2))^s (2G_\phi(\tau_1,\tau_2))^r. \nn
\end{align}
We will consider the bi-partition $A:B$ of the SYK cluster $\psi_j$, whose correlation are
\begin{align}
    G_A(\tau_1,\tau_2)=\frac1{N_A}\sum_{j=1}^{N_A} \psi_j(\tau_1)\psi_j(\tau_2),\quad 
    G_B(\tau_1,\tau_2)=\frac1{N_B}\sum_{j=N_A+1}^{N_B-N_A} \psi_j(\tau_1)\psi_j(\tau_2)
\end{align}
with $N=N_A+N_B$. Then the $-\S_\psi$ becomes
\begin{align}\begin{split}
    -\S_\psi=&\frac{N_A}2\log\det(\partial-\Sigma_A)+\frac{N_B}2\log\det(\partial-\Sigma_B)\\
    &~+\iint\kd{-\frac{N_A}2\Sigma_A G_A
    -\frac{N_B}2\Sigma_B G_B +N\frac{J^2}{4q}(2G_\psi)^q},
\end{split}\end{align}
where $G_\psi=\frac{N_A}{N}G_A+\frac{N_B}{N}G_B$. Here and after, we use the abbreviation $\iint=\int_0^\beta d\tau_1d\tau_2$ and omit the time dependence $(\tau_1,\tau_2)$ in $G$'s and $\Sigma$'s.

\subsection{R\'enyi mutual information}

To proceed, we set the initial state to be a thermofield double (TFD) state by doubling the Hilbert space to $\H_L\otimes\H_R$.
Explicitly, we consider a double-copy of the composite system $\H_L\otimes\H_R$ and define the maximally entangled state by
\begin{align}
    (\psi_j^L+i\psi_j^R)\ket0=(\phi_j^L+i\phi_j^R)\ket0=0, \ \forall j
\end{align}
{and} the TFD state is prepared by $\ket{\beta}=\frac1{\sqrt Z_\beta}e^{-\beta H_L/2}\ket0$. 
We can evolve the state $\ket\beta$ along time $t$ with the Hamiltonian $\L=H\otimes 1$ and reach a state $\ket{\beta+it}=e^{it\L}\ket\beta$. 

We will calculate the $n$-th R\'enyi mutual information between the subsystem $A=A_L\cup A_R$ and $B=B_L\cup B_R$, which contain the $n$-th R\'enyi entropies of $A$, $B$ and $A\cup B$ defined via replica trick
\begin{align}
    S_n(\mathcal R)=\frac{\log \Tr \rho(\mathcal R)^n}{1-n}
    =\frac{\log Z_n(\mathcal R)-n\log Z_1}{1-n}
    =\frac{\S_n(\mathcal R)-n\S_1}{n-1}, 
    \quad \mathcal R=A,\ B,\ A\cup B
\end{align}
where $\rho(\mathcal R)=\Tr_{\bar{\mathcal R}}\ket{\beta+it}\bra{\beta+it}$ is the reduced density matrix of the region $\mathcal R$. 
We have considered the large $N,M$ limit (with their ratio fixed) at the last step and $\S_n(\mathcal R)$ is the $n$ replica on-shell action with twist boundary condition on region $\mathcal R$.
To calculate the replica partition function $Z_n(\mathcal R)$, we replicate the Majorana as
$\psi_j^a,\phi_j^b$ with replica indexes $a,b=1,2,\cdots, n$. On replicated bi-local fields are
\begin{align}
    G_A^{ab}(\tau_1,\tau_2)=&\frac1{N_A}\sum_{j=1}^{N_A} \psi_j^a(\tau_1)\psi_j^b(\tau_2),\nn\\ 
    G_B^{ab}(\tau_1,\tau_2)=&\frac1{N_B}\sum_{j=N_A+1}^{N_B-N_A} \psi_j^a(\tau_1)\psi_j^b(\tau_2),\\ G_\phi^{ab}(\phi_1,\phi_2)=&\frac1{M}\sum_{j=1}^{M} \phi_j^a(\tau_1)\phi_j^b(\tau_2).\nn
\end{align}
The replica actions are
\begin{align}\label{SYKReplicaAction}
    \S_n=&\S_{\psi,n}+\S_{\phi,n}+\S_{i,n},\\ 
    -\S_{\psi,n}=&N\frac12\log\det(\partial-\Sigma_\psi^{ab})+N\iint \sum_{ab} \kd{-\frac12\Sigma_\psi^{ab} G_\psi^{ab}+\frac{J^2}{4q}(2G_\psi^{ab})^q}, \nn\\
    -\S_{\phi,n}=&M\frac12\log\det(\partial-\Sigma_\phi^{ab})+M\iint \sum_{ab} \kd{-\frac12\Sigma_\phi^{ab} G_\phi^{ab}+\frac{K^2}{4p}
(2G_\phi^{ab})^p}, \nn\\
    -\S_{i,n}=&N\iint \sum_{ab}\frac{V^2}{4s}(2G_\psi^{ab})^s (2G_\phi^{ab})^r. \nn
\end{align}
We impose the twist boundary conditions 
\footnote{
Although our method is equivalent to that in \cite{Penington:2019kki,Gu:2017njx}, the time parametrization and the replica indexes here are different, so that only the contours with the same replica index are coupled by interaction terms.}
\begin{align}\label{eq:ReplicaBC}
    \psi_j^a(\tau_*^-)=\psi_j^{a+1}(\tau_*^+),\quad 
    \psi_j^a(0^+)=\psi_j^{a+1}(0^-),\quad 
\end{align}
where $\psi_j^{n+1}=\psi_j^1$, for the $\psi_j^a$ with index $j$ depending on the entanglement regions
\begin{align}\begin{split}
    S_n(A): \ &\quad  1\leq j\leq N_A,\\
    S_n(B): \ &\quad  N_A+1\leq j\leq N,\\
    S_n(A\cup B): \ &\quad  1\leq j\leq N.
\end{split}\end{align} 
We will consider the analytic continuation from $\tau_*$ to $\beta+it$. 
Also, we should apply the ordinary anti-periodic boundary condition for each fermion loop, which depends on the twist boundary conditions.

Without the interactions, namely $J=K=V=0$, the untwist solution, and the twist solution are respectively {given by}
\begin{align}\label{SYKFreeTwist}
    G_0^{ab}(\tau_1,\tau_2)=\frac12\sgn(\tau_1-\tau_2)\delta_{ab},\quad
    G_0^{ab,T}(\tau_1,\tau_2)=G_0^{a-\theta_*(\tau_1),b-\theta_*(\tau_2)}(\tau_1,\tau_2),
\end{align}
where $\theta_*(\tau)=\Theta(\tau-\tau_*)\Theta(\beta-\tau_*-\tau)$ and $\Theta(\tau)$ is the step function. They are {free} propagators subjected to the boundary conditions. We will use them to define the untwist and twist differential operators.

\begin{figure}
    \centering
    \includegraphics[height=0.4\linewidth]{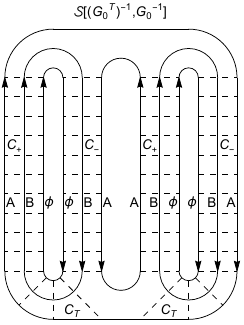}~~
    \includegraphics[height=0.4\linewidth]{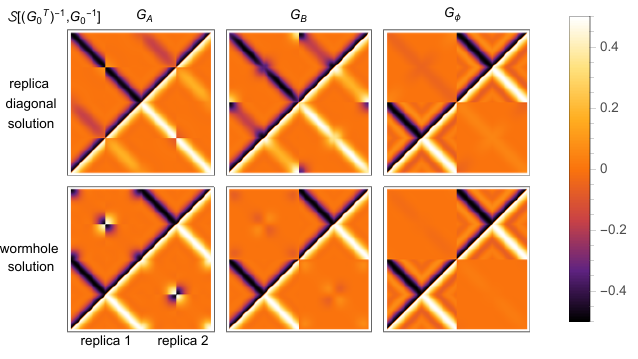}
    \caption{Contours and solutions for second-replicated action $\S[(G_0^T)^{-1},G_0^{-1}]$. The arrows in the contours denote the time direction and the dashed lines denote the interactions. The parameters for the solutions are $J=K=1,\,V=0.5,\,q=4,\,p=2,\,s=r=1,\,N/M=0.1,\, N_A/N=0.75,\,\beta=0$, and $t=2$.}
    \label{fig:SYKSolution}
\end{figure}

For numerical convenience, we can equivalently work in the time domain $(0,n\beta)$ and identify
$\psi_j(\tau+a\beta)=\psi_j^a(\tau)$. To take the twist boundary condition into account, we introduce two undetermined differential operators $X,Y$ for subsystems $A$ and $B$ respectively, and rewrite the replicated action as
\begin{align}
    -\S[X,Y]=&-\S_\phi-\S_i
    +\frac{N_A}2\log\det(X-\Sigma_A)+\frac{N_B}2\log\det(Y-\Sigma_B)\\
    &+\iint_C \kd{-\frac{N_A}2\Sigma_A G_A-\frac{N_B}2\Sigma_B G_B+N\frac{J^2}{4q}f(\tau_1)f(\tau_2)(2G_\psi)^q},\nn
\end{align}
where the integral $\iint_C=\int_{C} d\tau_1d\tau_2$ goes along the contour $C=C_T\cup C_+\cup C_-$ with thermal contour $C_T=\cup_{a=1}^{n}((a-1)\beta^+,a\beta^-)$, forward contour $C_+=\cup_{a=1}^{n}(a\beta^-,a\beta+it)$ and backward contour $C_-=\cup_{a=1}^{n}(a\beta+it,a\beta^+)$. 
The entire contour $C$ for $n=2$ with a twist on subsystem $A$ is illustrated in the left panel of Fig.~\ref{fig:SYKSolution}. 
To take into account the direction of time, we have inserted functions $f(\tau_1)f(\tau_2)$ into the interaction terms in $\S_\psi,\, \S_\phi, \S_i$ where $f(\tau)=1,i,-i$ for $\tau\in C_T,C_+,C_-$ respectively.
The differential operators $X,Y$ will take the value of either $G_{0}^{-1}$ or $(G_{0}^{T})^{-1}$ with propagators $G_{0}(\tau_1+a\beta,\tau_2+b\beta)=G_0^{ab}(\tau_1,\tau_2),\, G_{0}^T(\tau_1+a\beta,\tau_2+b\beta)=G_0^{ab,T}(\tau_1,\tau_2)$ defined in  \eqref{SYKFreeTwist}.

In the large $N,M$ limit, the Schwinger-Dyson (SD) equation reads
\begin{align}\begin{split}
    &G_A^{-1}=X-\Sigma_A,\quad 
    G_B^{-1}=Y-\Sigma_B,\quad 
    G_\phi^{-1}=G_0^{-1}-\Sigma_\phi,\\
    &\Sigma_A=\Sigma_B=f(\tau_1)f(\tau_2)\kd{J^2 (2G_\psi)^{q-1}
    +V^2 (2G_\psi)^{n-1}(2G_\phi)^m},\\
    &
    \Sigma_\phi=f(\tau_1)f(\tau_2)\kd{K^2 (2G_\phi)^{p-1}+\frac {rN}{sM}V^2(2G_\psi)^s(2G_\phi)^{r-1}}.
\end{split}\end{align}
Using the solution of the SD equation with given differential operators $X,Y$, we can write the on-shell action as
\begin{align}
    \S[X,Y]=&~-\frac{N_A}2\log\det(G_A^{-1})
    -\frac{N_B}2\log\det(G_B^{-1})
    -\frac{M}2\log\det(G_\phi^{-1})\nn\\
    &~+\iint_C f(\tau_1)f(\tau_2)\Big[N\kc{1-\frac1q}\frac{J^2}{4}(2G_\psi)^q
    +M\kc{1-\frac1p}\frac{K^2}{4}(2G_\phi)^p\\
    &~+N\kc{1+\frac rs-\frac1s}\frac{V^2}{4}(2G_\psi)^s(2G_\phi)^r \Big].\nn
\end{align} 
Finally, the R\'enyi entropies and mutual information are given by
\begin{align}\begin{split}\label{SYKEntropies}
    (n-1)S_n(A)=&~\S[(G_0^T)^{-1},G_0^{-1}]-\S_0,\\
    (n-1)S_n(B)=&~\S[G_0^{-1},(G_0^T)^{-1}]-\S_0,\\
    (n-1)S_n(A\cup B)=&~\S[(G_0^T)^{-1},(G_0^T)^{-1}]-\S_0,\\
    (n-1)I_n(A:B)=&~\S[(G_0^T)^{-1},G_0^{-1}]+\S[G_0^{-1},(G_0^T)^{-1}]
    -\S[(G_0^T)^{-1},(G_0^T)^{-1}]-\S_0,
\end{split}\end{align}
where $\S_0=\S[G_0^{-1},G_0^{-1}]$.

\begin{figure}
    \centering
    \includegraphics[height=0.28\linewidth]{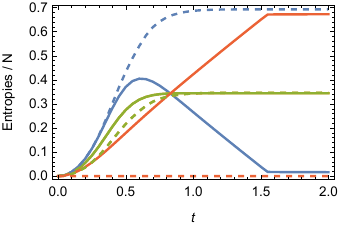}
    \includegraphics[height=0.28\linewidth]{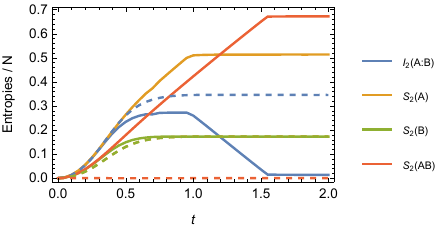}
    \caption{Renyi mutual information $I_2(A:B)$ and Renyi entropies $S_2(A),\ S_2(B),\ S_2(AB)$ in the SYK cluster coupled to bath. The parameters are $J=K=1$, (solid) $V=0.5$ or (dashed) $V=0$, $q=4,\,p=2,\,s=r=1,\,N/M=0.1,\, \beta=0$, and (left) $N_A/N=0.5$ or (right) $N_A/N=0.75$.}
    \label{fig:SYKMI}
\end{figure}

\begin{figure}
    \centering
    \includegraphics[height=0.28\linewidth]{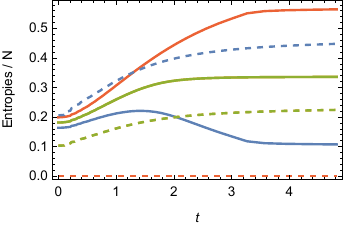}
    \includegraphics[height=0.28\linewidth]{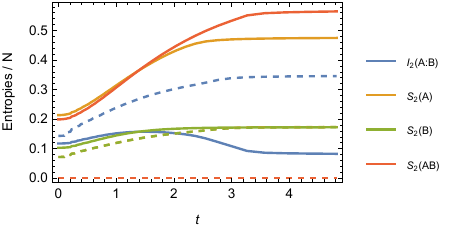}
    \caption{Renyi mutual information $I_2(A:B)$ and Renyi entropies $S_2(A),\ S_2(B),\ S_2(AB)$ in the SYK cluster coupled to bath. The parameters are the same as that in Fig.~\ref{fig:SYKMIbeta}, except $\beta=2$.}
    \label{fig:SYKMIbeta}
\end{figure}

We will numerically solve the SD equation for the $n=2$ replica. 
At the late time, there are two solutions in the replica action, including one replica diagonal solution and one wormhole solution, similar to \cite{Penington:2019kki}. 
In Fig.~\ref{fig:SYKSolution}, we show the contours and two solutions for action $\S[(G_0^T)^{-1},G_0^{-1}]$ at the late time.  
From these two solutions, we should choose the one that results in a smaller value of the on-shell action. 
Finally, substituting the on-shell actions into \eqref{SYKEntropies} for different time $t$, we obtain the time evolution of R\'enyi entropy and R\'enyi mutual information, as shown in Figs.~\ref{fig:SYKMI} and \ref{fig:SYKMIbeta}. 
We mainly focus on the case of $N_B\leq N_A\ll M$, $q=4,\, p=2,\, s=r=1$ and $V\lesssim J\sim K$. 
So the SYK cluster itself is chaotic and the bath itself is integrable. 
In general, the {entropy} and mutual information {undergo} through four stages:
\begin{enumerate}
    \item All orders of R\'enyi entropy and the mutual information increase due to the interactions.
    \item The entropy $S_2(B)$ is saturated. 
    The growth of $S_2(A),S_2(AB)$ slightly slows down. 
    The mutual information $I_2(A:B)$ stops growing.
    \item The entropy $S_2(A)$ is saturated. 
    The growth of $S_2(AB)$ slightly slows down. 
    The mutual information $I_2(A:B)$ decays.
    \item The entropy $S_2(AB)$ is saturated. 
    The mutual information $I_2(A:B)$ reaches a plateau. 
\end{enumerate}
Notice that both the initial value and the final value of the mutual information $I_2(A:B)$ are zero for $\beta=0$ and finite for $\beta>0$. 
The evolution of the mutual information $I_2(A:B)$, indicates the entangling between $A$ and $B$ due to the $J$ interaction and disentangling due to the thermalization from the bath via the $V$ interaction. 

To study the effect of baths on the internal entanglement of the SYK clusters, we compare the evolution of mutual information and entropies with and without the interaction between the SYK clusters and the baths in Figs.~\ref{fig:SYKMI} and \ref{fig:SYKMIbeta}. As expected, $S_2(A:B)=0$ always and $S_2(A),\,S_2(B),\,I_2(A:B)$ grow and then become saturated. We can compare the case of $V\neq0$ and $V=0$ to see the impact of baths. 

When $\beta=0$, we observe that the entropies $S_2(A)$ and $S_2(B)$ in the $V\neq0$ case grow faster than that in the $V=0$ case before their saturation. The enhancement of the growth rate is a natural consequence of the additional thermalization from baths. While the mutual information in the $V\neq0$ case grows at the same rate as that in the $V=0$ case before the saturation of either $S_2(A)$ or $S_2(B)$. It implies that the thermalization due to the baths hardly 
affects the entanglement between $A$ and $B$ until one of their entropies becomes saturated.

When $\beta>0$, all the growth of $S_2(A),\,S_2(B),\,I_2(A:B)$ are affected by the bath. Their growth rates in the $V=0$ case are different from {those} in the $V\neq 0$ case. The reason is that the baths already affect the preparation of the initial state. In other words, the TFD state is given by the combined Hamiltonian \eqref{eq:SYKHam} including the interaction term with the baths. It is different from the initial state at $\beta=0$, which is a maximally entangled state and is independent of the Hamiltonian.


\section{Conclusion and outlook}\label{sec:conc}
In this paper, we have investigated the dynamic process of interior entanglement of two strongly coupled systems, including the gravity and SYK systems. 
The first model we have proposed is a doubly holographic model in $(3+1)$-dimensions, where the CFT matter sectors are dual to a $4$-dimensional bulk and the black hole is modeled by an EOW brane in the bulk. 
In the zero tension limit, the bulk geometry can be described by the standard SAdS metric with Neumann boundary conditions on the branes. 
By introducing a circular compactification in the $y$ direction, we have obtained a finite-sized black hole system from the boundary perspective. 
In this scenario, we have constructed the EWCS and the two-dimensional inhomogeneous HRT surface to quantify the reflected entropy and mutual information between the black hole bipartition $\mathcal{BH}_1$ and $\mathcal{BH}_2$, respectively.
Generally, the reflected entropy and mutual information both increase increase at early times due to the internal interactions of the black hole system. 
While at late times, these interior entanglements seep into the environment, since the interactions between the black hole system and the radiation drive the system to the final equilibrium. 
Moreover, we have extracted two universal patterns of evolution: 
For large black holes, no plateau emerges during the evolution, and the reflected entropy directly declines from the maximum to the final equilibrium value. 
For small black holes, a plateau emerges during the evolution of the reflected entropy, before the final equilibrium at a lower value.
Similarly for mutual information, a corresponding ramp-slope-stabilizing exists in the dynamical evolution.

Our second model introduces a joint system comprising a SYK cluster, denoted as $\psi_j$, coupled to a large SYK bath, $\phi_j$. 
Specifically, the DOF in {both SYK} clusters and the bath are defined by the number of fermions they contain, expressed as $N= N_A + N_B$ and $M$, with $A$ and $B$ representing the bi-partition of the SYK clusters. 
In the large $N$ limit, this system is analyzed by numerically solving the SD equations. 
We have computed the second R\'enyi entropies and mutual information using the replica trick in this joint model. 
Our study covers different-sized bi-partitions of the SYK cluster, with $N_{A} \leq N_{B}$.
Generally, we observe an initial increase in the mutual information of the bi-partition due to the internal interactions within the SYK cluster. 
At late times, the entanglement entropies of the subsystems saturate, while the entanglement between the SYK cluster and the bath continues to evolve. 
Consequently, the interior entanglement gradually leaks into the bath, eventually stabilizing at a relatively low equilibrium level.

For the first model, the zero tension condition is usually imposed in literature to circumvent the backreaction of the brane to the ambient geometry in the investigation of the dynamical evolution \cite{Ling:2020laa,Liu:2022pan}. 
Therefore, an intriguing area for future research lies in examining the dynamics within a backreacted spacetime. 
This challenge could potentially be addressed by employing the ingoing Eddington coordinates within the Einstein-DeTurck formulation \cite{Figueras:2012rb}. 
It is important to note that in such a framework, the Einstein equations are transformed into a mixed elliptic-hyperbolic system, raising issues related to the local uniqueness of solutions \cite{Dias:2015nua}. 
Aside from stationary scenarios, another promising avenue involves extending the numerical setup to {non-equilibrium} cases \cite{Janik:2017ykj,Chen:2019uhq,Chen:2022cwi,Chen:2022tfy,Zhang:2021etr,Zhang:2022cmu,Chen:2022vag,Chen:2023eru}, expanding the scope and applicability of the study to more dynamic and complex systems.

In both the gravity and SYK models, we have noticed a pronounced difference between the evolution patterns of the reflected entropy and the mutual information. 
For any holographic state, there often exists a substantial gap of order $N^2$ between these two measures, known as the Markov gap \cite{Hayden:2021gno}. 
This gap is commonly associated with the fidelity of a specific Markov recovery problem and has been a topic of extensive exploration in the context of double holography \cite{Lu:2022cgq,Afrasiar:2022fid,Afrasiar:2023jrj,Basak:2023bnc}. 
By comparing the behaviors of the reflected entropy and the mutual information, we anticipate a noticeable enhancement of the Markov gap at late times but before the Page time. The reason for this enhancement deserves investigation in future studies.

Finally, in the model of SYK clusters coupled to baths, we can approximate the dynamics of the SYK clusters using the Lindblad master equation in the Born-Markov approximation \cite{Kawabata:2022osw, Kulkarni:2021gtt, Garcia-Garcia:2022adg}. This approximation is expected to be valid under the conditions $N \ll M$ and $V \ll J \ll K$. The slope-stabilizing behavior observed in the mutual information here is akin to the slope-stabilizing behavior observed in the purity in the Lindbladian SYK \cite{Wang:2021woy}, although in our case, the bipartite systems $A$ and $B$ are entangled due to their interactions during time evolution, rather than being initially configured to be in a TFD state. 
It would still be interesting to apply a large $q$ analysis as well as a quantum trajectories analysis in our model.


\section*{Acknowledgments}
We are grateful to Pritam Chattopadhyay, Sayantan Choudhury, and Yuan Sun for helpful discussions. Yuxuan Liu is supported by the Beijing Natural Science Foundation under Grant No. 1222031 and the National Natural Science Foundation of China under Grant No.~12035016. He is particularly grateful to his wife Peiwen Cao. 
Shao-Kai Jian is supported by a startup fund at Tulane University. 
Yi Ling is supported in part by the Beijing Natural Science Foundation under Grant No. 1222031, and the Innovative Projects of Science and Technology at IHEP. He is also supported by the National Natural Science Foundation of China under Grant No.~12035016 and 12275275. Zhuo-Yu Xian is funded by DFG through the Collaborative Research Center SFB 1170 ToCoTronics, Project-ID 258499086—SFB 1170, as well as by Germany's Excellence Strategy through the W\"urzburg‐Dresden Cluster of Excellence on Complexity and Topology in Quantum Matter ‐ ct.qmat (EXC 2147, project‐id 390858490). Zhuo-Yu Xian also acknowledges support from the National Natural Science Foundation of China under Grant No.~12075298.

\bibliographystyle{JHEP}
\bibliography{refs}
	
\end{document}